\batchmode
\makeatletter
\def\input@path{{"C:/Users/saied/AKBARIS/Saia/Papers/[2025-ACC, CSL] [In Progress] - Lb Stochastic DNN/"}}
\makeatother
\documentclass[english]{IEEEtran}
\usepackage[T1]{fontenc}
\usepackage[latin9]{inputenc}
\usepackage{color}
\usepackage{amsmath}
\usepackage{amsthm}
\usepackage{amssymb}
\usepackage{stackrel}
\usepackage{graphicx}
\usepackage{esint}

\makeatletter
\theoremstyle{definition}
\newtheorem{assumption}{Assumption}
\theoremstyle{definition}
\newtheorem{defn}{\protect\definitionname}
\theoremstyle{plain}
\newtheorem{lem}{\protect\lemmaname}
\newenvironment{lyxlist}[1]
	{\begin{list}{}
		{\settowidth{\labelwidth}{#1}
		 \setlength{\leftmargin}{\labelwidth}
		 \addtolength{\leftmargin}{\labelsep}
		 }}
	{\end{list}}
\theoremstyle{plain}
\newtheorem{thm}{\protect\theoremname}

\usepackage{cite}
\usepackage{multirow}
\usepackage{tikz}
\usetikzlibrary{decorations.pathmorphing, arrows.meta}

\makeatother

\usepackage{babel}
\providecommand{\definitionname}{Definition}
\providecommand{\lemmaname}{Lemma}
\providecommand{\theoremname}{Theorem}

\begin{document}
\title{Lyapunov-Based Deep Neural Networks for Adaptive Control of Stochastic
Nonlinear Systems}
\author{Saiedeh Akbari, Cristian F. Nino, Omkar Sudhir Patil, Warren E. Dixon$^{*}$\thanks{$^{*}$Saiedeh Akbari, Cristian F. Nino, Omkar Sudhir Patil, and Warren
E. Dixon are with the department of Mechanical Engineering and Aerospace
Engineering, University of Florida, Gainesville, FL, 32611-6250 USA.
Email: \{akbaris, cristian1928, patilomkarsudhir, wdixon\}@ufl.edu. }\thanks{This research is supported in part by AFOSR grant FA9550-19-1-0169.
Any opinions, findings, and conclusions or recommendations expressed
in this material are those of the author(s) and do not necessarily
reflect the views of the sponsoring agencies.}}
\maketitle
\begin{abstract}
Controlling nonlinear stochastic dynamical systems involves substantial
challenges when the dynamics contain unknown and unstructured nonlinear
state-dependent terms. For such complex systems, deep neural networks
can serve as powerful black box approximators for the unknown drift
and diffusion processes. Recent developments construct Lyapunov-based
deep neural network (Lb-DNN) controllers to compensate for deterministic
uncertainties using adaptive weight update laws derived from a Lyapunov-based
analysis based on insights from the compositional structure of the
DNN architecture. However, these Lb-DNN controllers do not account
for non-deterministic uncertainties. This paper develops Lb-DNNs to
adaptively compensate for both the drift and diffusion uncertainties
of nonlinear stochastic dynamic systems. Through a Lyapunov-based
stability analysis, a DNN-based approximation and corresponding DNN
weight adaptation laws are constructed to eliminate the unknown state-dependent
terms resulting from the nonlinear diffusion and drift processes.
The tracking error is shown to be uniformly ultimately bounded in
probability. Simulations are performed on a nonlinear stochastic dynamical
system to show efficacy of the proposed method.
\end{abstract}


\global\long\def\SS{\mathbb{S}}%
\global\long\def\RR{\mathbb{R}}%
\global\long\def\nz{\left\Vert z\right\Vert }%
\global\long\def\ne{\left\Vert e\right\Vert }%
\global\long\def\nt{\left\Vert \widetilde{\theta}\right\Vert }%
\global\long\def\nzz{\left\Vert z\right\Vert ^{2}}%
\global\long\def\nee{\left\Vert e\right\Vert ^{2}}%
\global\long\def\ntt{\left\Vert \widetilde{\theta}\right\Vert ^{2}}%
\global\long\def\tq{\triangleq}%
\global\long\def\Linf{\mathcal{L}_{\infty}}%
\global\long\def\yy{\mathcal{Y}}%
\global\long\def\uu{\mathcal{U}}%
\global\long\def\grad{\mathcal{\nabla_{\widetilde{\theta}}}}%
\global\long\def\tvec{\mathcal{\text{vec}}}%
\global\long\def\sgn{\mathcal{\text{sgn}}}%
\global\long\def\nsig{\left\Vert \Sigma\Sigma^{\top}\right\Vert _{\infty}}%

\section{Introduction\label{sec: Introduction}}

Stochastic control theory has received considerable attention, due
to its far-reaching applicability across various domains \cite{Deng.Krstic1997,Deng.Krstic.ea2001,Jiang.Karimi.ea2020,Stojanovic.Nedic2016,Liu.Zhang.ea2007,Li.Krstic2022}.
Modeling the uncertainty in the system as a stochastic process can
make the control design less conservative than assuming the worst-case
upper bounds of the uncertainty \cite{Kumar.Varaiya2015}. However,
due to the stochastic nature of such systems, many challenges are
introduced in designing an adaptive controller for such systems.

To illustrate the challenges involved in designing an adaptive controller
for stochastic systems, consider a nonlinear stochastic system of
the form
\begin{equation}
{\rm d}x=F\left(x,u\right){\rm d}t+G\left(x,t\right){\rm d}\omega,\label{eq: general SDE}
\end{equation}
where $x$ denotes the state, $F$ denotes the drift process, and
$G$ denotes the diffusion process. For many practical systems, the
drift and diffusion processes may be expressed in a control-affine
form given by $F\left(x,u\right)\tq f\left(x\right)+g_{1}\left(x\right)u$
and $G\left(x,t\right)\tq g_{2}\left(x\right)\Sigma\left(t\right)$,
where $f$, $g_{1}$, and $u$ respectively denote to drift vector
field, control effectiveness, and the control input, and $g_{2}$
is a matrix-valued diffusion function and $\Sigma$ is a nonnegative
matrix-valued function. To address the control problem for systems
modeled as (\ref{eq: general SDE}), early works in the literature
assume model knowledge of $F$ and $G$ (e.g., \cite{Deng.Krstic1997,Deng.Krstic.ea2001}).
Although several results compensate for drift uncertainties, fewer
works pay attention to compensating for diffusion uncertainties.

To compensate for the drift and diffusion uncertainties, different
robust and adaptive compensation approaches have been developed (e.g.,
\cite{Zhou.Shi.ea2012,Li.Krstic2020,Shi.Xia.ea2006,Qi.Zong.ea2019,Chen.Liu.ea2013,Li.Sui.ea2016,Sun.Chen.ea2016,Li.Chen.ea2009}).
Shallow NNs are often used to compensate for uncertainties in nonlinear
stochastic systems due to their inherent function approximation capabilities,
to fulfill either stabilization or trajectory tracking objectives
(e.g., \cite{Li.Chen.ea2009,Chen.Liu.ea2013,Li.Li.ea2018,Wang.Liu.ea2019a,Wang.Chen2020,Wang.You.ea2022,Chen.Jiao.ea2009}).
Despite considerable advances, controlling nonlinear stochastic systems
involves enduring.
\begin{itemize}
\item Assumption of known sign/value of bounds or known structure of the
uncertainties: In results such as \cite{Wang.Chen2020} and \cite{Wang.You.ea2022},
the NN is only used to approximate $f\left(x\right)$, whereas $g_{2}\left(x\right)$
is upper bounded. Additionally, in results such as \cite{Li.Chen.ea2009}
and \cite{Chen.Jiao.ea2009}, diffusion and drift uncertainties are
assumed have a known linear-in-parameter structure, where the unknown
parameters are estimated using an adaptive approach. For example,
\cite{Chen.Liu.ea2013} requires a bound with known sign on the drift
vector. In this paper, assumptions on structure of the drift and diffusion
uncertainties and any knowledge on the value or sign of their bounds
are avoided. 
\item No probabilistic analysis for the escape risk: Studies such as \cite{Chen.Liu.ea2013,Li.Li.ea2018,Wang.Liu.ea2019a,Wang.Chen2020,Wang.You.ea2022}
do not provide probabilistic analysis to account for the risk of the
states escaping from the compact set, where the universal function
approximation property is guaranteed to hold. For stochastic dynamics,
there is a probability of solutions escaping the compact set, even
when initialized within the compact set \cite[Chap. 3]{Kushner1971}.
Even for deterministic systems, the state trajectories need to be
shown to lie in a compact set by appropriately shaping the set \cite[Def. 4.6]{Khalil2002}.
A unique contribution of this paper is a lemma that expresses the
probability of escape in terms of the ultimate bound and the initial
condition of the Lyapunov function.
\item Vanishing stochastic noise: A key assumption across studies with global
asymptotic stable (GAS) results (such as \cite{Deng.Krstic1997,Deng.Krstic.ea2001})
and uniformly ultimately bounded (UUB) results (such as \cite{Li.Krstic2022,Wang.Liu.ea2019a,Wang.Chen2020,Li.Chen.ea2009,Chen.Jiao.ea2009,Chen.Liu.ea2013}),
is that the equilibrium is preserved under the presence of the noise,
i.e., $g_{2}\left(0\right)=0$. This assumption is conservative since
it implies that once perfect tracking is achieved and the system is
at equilibrium, the stochastic noise disappears. Such an assumption
is not applicable to most real-world applications and is avoided in
this paper. Furthermore, a correlation between the noise at the origin
and the ultimate bound is shown that offers an insightful examination
of the escape risk.
\end{itemize}
Deep neural networks (DNNs) leverage the nested nonlinearities arising
from function composition, resulting in enhanced function approximation
performance even with fewer overall parameters when compared to a
corresponding shallow NN, as evidenced by empirical and experimental
findings \cite{Rolnick.Tegmark2018,Liang.Srikant2016,Lamb.Bell.ea2023}.
Most existing DNN results use offline training by using sampled input-output
datasets (e.g., \cite[Sec. 6.6]{Brunton.Kutz2019} and \cite{Noda.Arie.ea2014,Sarikaya.Corso.ea2017,Nguyen.Cheah2022}).
The aforementioned DNNs are applied as feedforward terms, and once
implemented, the weights of the DNNs cannot be updated in real-time
via a Lyapunov stability driven update law. Hence, this open-loop
implementation lacks stability guarantees. Recent breakthroughs address
the lack of stability and adaptation to unexpected uncertainties by
introducing Lyapunov-based DNNs (Lb-DNN) which use Lyapunov-based
techniques to update the weights of the DNN in real-time \cite{Patil.Le.ea2022,Patil.Le.ea.2022,Hart.patil.ea2023,Nino.Patil.ea2023,Griffis.Patil.ea.2023}.
Specifically, the weights of Lb-DNNs are adjusted from analytic update
laws designed from a Lyapunov-based stability analysis, allowing for
real-time weight updates of the DNN without requiring pre-training
from offline data \cite{Patil.Le.ea2022,Patil.Le.ea.2022}. To the
best of the authors' knowledge, the result in \cite{Chen.Mei.ea2024}
is only one result that employs Lb-DNNs for control of nonlinear stochastic
systems. In \cite{Chen.Mei.ea2024}, Lb-DNNs are used for an optimized
backstepping control design in a class of stochastic nonlinear strict-feedback
systems. However, the probability of solutions escaping the compact
set is not considered and the assumption of noise vanishing at the
equilibrium is made.

In this paper, the trajectory tracking problem is addressed for a
general class of control-affine nonlinear stochastic dynamical systems
with unstructured uncertainties. To allow for targeted adjustments
for each uncertain component (i.e., terms stemming from $f\left(x\right)$,
$g\left(x\right)$, and $\Sigma\left(t\right)$), the developed method
uses three Lb-DNNs to approximate the inherent deterministic and stochastic
uncertainties within the closed-loop error system.\footnote{One DNN architecture may also be used instead of three to compensate
for the uncertainties, however, using multiple DNNs allows for targeted
adjustments.} The main contributions of this paper are summarized as below.
\begin{enumerate}
\item Lb-DNNs are utilized to compensate for the uncertainties of a nonlinear
stochastic dynamical system. By taking advantage of the superior function
approximation capabilities of DNNs and improved performance of the
Lb-DNNs compared to shallow NNs \cite{Patil.Le.ea2022}, a multi-DNN
technique is developed such that three different Lb-DNNs individually
approximate components stemming from drift and diffusion uncertainties.
\item The assumption of vanishing noise is relaxed by a strategic use of
Taylor's theorem and properties of the trace operator in the subsequent
stability analysis. Relaxing this assumption allows for non-vanishing
noise, which adds to the challenges in the stability analysis, but
makes the developed method better suited for real-world applications.
Furthermore, an additional generality is added that the noise has
an unknown time-varying covariance matrix that is multiplied by the
nonlinearity of the diffusion matrix. Additionally, there are no assumptions
on the structure of the uncertainties. Moreover, any assumption regarding
knowledge on the uncertainties or the sign and value of their upper-bounds
are relaxed compared to the literature.
\item A Lyapunov-based stability analysis is conducted to guarantee probabilistic
exponential convergence to an ultimate bound about the origin and
the boundedness of all the signals. Through the stability analysis,
the tracking error is shown to be uniformly ultimately bounded in
probability and a theorem is developed to quantify the probability
of stability. For the universal approximation property to hold, all
states that are inputs to the DNN must be constrained to a compact
set, for all time. Due to the stochastic nature of the dynamics, the
stability result is proven in probability, meaning that there is a
risk that states may escape the aforementioned compact set. From a
practical perspective, quantifying the escape risk is essential to
provide a probabilistic certification on the designed controller.
Hence, it is crucial to quantify this risk via a probability analysis.
Although for asymptotically stable results, a probability analysis
is typically provided (see \cite{Deng.Krstic1997,Li.Chen.ea2009,Chen.Jiao.ea2009}),
for results with uniform ultimate bounds in probability, a probability
analysis is usually not provided.
\end{enumerate}
Simulations are performed on a five-dimensional nonlinear stochastic
dynamical system to show the efficacy of the proposed method. Additional
simulations illustrate the tracking perfomance of the proposed controller
in response to variations in mean and covariance of the stochastic
noise. 

\section{Notation\label{sec: Notation}}

Let $\mathbf{0}_{m\times n}$ and $I_{m\times n}$ denote the $m\times n$
dimensional zero and identity matrices, respectively. The universal
quantifier $\forall\varpi\left(x\right)$ asserts that $\varpi\left(x\right)$
is true for every $x$ in the domain. The existential quantifier $\exists x$
$\varpi\left(x\right)$ asserts that $\varpi\left(x\right)$ is true
for at least one $x$ in the domain. Implication $\varpi\implies\varphi$
is defined as $\lnot\varpi\vee\varphi$, where negation $\lnot\varpi$
means $\varpi$ is false, and disjunction $\varpi\vee\varphi$ means
at least one of $\varpi$ or $\varphi$ is true. The colon symbol
$:$ in set-builder notation, as in $\left\{ x:\varpi\left(x\right)\right\} $,
means ``such that'' and specifies the element $x$ in the domain
for which the condition $\varpi\left(x\right)$ holds. The right pseudo
inverse of a full-row-rank matrix $A$ is defined as $A^{+}\left(\cdot\right)\tq A^{\top}\left(\cdot\right)\left(A\left(\cdot\right)A^{\top}\left(\cdot\right)\right)^{-1}$.
The expected value is given by ${\rm E}\left[X\right]=\intop_{-\infty}^{\infty}x\cdot\mathbf{F}\left(x\right){\rm d}x$,
where $\mathbf{F}\left(x\right)$ is the probability density function
of the continuous random variable $X$. For a matrix $A\tq\left[a_{i,j}\right]\in\RR^{n\times m}$,
where $a_{i,j}$ is the element on the $i^{\text{th}}$ row of the
$j^{\text{th}}$ column of the matrix, the vectorization operator
is defined as $\tvec\left(A\right)=\left[a_{1,1},\ldots,a_{n,1},\ldots,a_{1,m},\ldots,a_{n,m}\right]^{\top}\in\RR^{nm}$.
For a square matrix $A\in\RR^{n\times n}$, the trace operator is
defined as $\text{tr}\left(A\right)=\stackrel[i=1]{n}{\sum}a_{i,i}$,
where $a_{i,i}$ represents the element on the $i^{\text{th}}$ row
of the $i^{\text{th}}$ column. From \cite[Chapter 2, Eq. 13]{Magnus.Neudecker2019},
the trace to vector property 
\begin{equation}
\text{tr}\left(A^{\top}B\right)=\tvec\left(A\right)^{\top}\tvec\left(B\right)\label{eq: trace to vec}
\end{equation}
holds for matrices $A$ and $B$. From \cite[Chapter 1, Eq. 25]{Magnus.Neudecker2019},
\begin{equation}
\text{tr}\left(ABC\right)=\text{tr}\left(BCA\right)=\text{tr}\left(CAB\right).\label{eq: order of multiplication property}
\end{equation}
Additionally, if $A$ and $B$ are positive semi-definite matrices,
then 
\begin{equation}
\text{tr}\left(AB\right)\leq\text{tr}\left(A\right)\text{tr}\left(B\right).\label{eq: separating trace inequality}
\end{equation}
The right-to-left matrix product operator is represented by $\stackrel{\curvearrowleft}{\prod}$,
i.e., $\stackrel{\curvearrowleft}{\stackrel[p=1]{m}{\prod}}A_{p}=A_{m}\ldots A_{2}A_{1}$
and $\stackrel{\curvearrowleft}{\stackrel[p=a]{m}{\prod}}A_{p}=I$
if $a>m$. The Kronecker product is denoted by $\otimes$. From \cite{Patil.Le.ea.2022}
and given any $A\in\mathbb{R}^{p\times a}$, $B\in\mathbb{R}^{a\times r}$,
and $C\in\mathbb{R}^{r\times s}$, differentiating $\mathrm{vec}\left(ABC\right)$
on both sides with respect to $\mathrm{vec}\left(B\right)$ yields
\begin{eqnarray}
\frac{\partial}{\partial\mathrm{vec}\left(B\right)}\mathrm{vec}\left(ABC\right) & = & C^{\top}\otimes A.\label{eq:vec_diff_prop}
\end{eqnarray}
Given a function $h:\RR^{n}\to\RR^{n}$, the notation $\underset{a\to b^{-}}{\lim}h\left(a\right)$
denotes the left-hand limit of $h$ at $b$. The $p$-norm is denoted
by $\left\Vert \cdot\right\Vert _{p}$, where the subscript is suppressed
when $p=2$. The Frobenius norm is denoted by $\left\Vert \cdot\right\Vert _{F}\triangleq\left\Vert \mathrm{vec}(\cdot)\right\Vert $.
For a bounded function $f:\RR_{\geq0}\to\RR^{n\times m}$, $\left\Vert f\right\Vert _{F\infty}\tq\underset{t\in\RR_{\geq0}}{\sup}\left\Vert f\right\Vert _{F}$.
The space of $k$-times differentiable functions is denoted by $\mathcal{C}^{k}$,
and a $\mathcal{C}^{\infty}$-smooth function is an infinitely differentiable
function. For $A\subseteq\mathbb{R}^{n}$ and $B\subseteq\mathbb{R}^{m}$,
let $C\left(A,B\right)$ denote the set of continuous functions $f:A\to\mathbb{R}^{m}$
such that $f\left(A\right)\subseteq B$, and let $C\left(A\right)\triangleq C\left(A,A\right)$.
In the filtered probability space of $\left({\bf \Omega},\,\mathbb{F},\,\mathbb{F}_{t},{\rm P}\right)$,
${\bf \Omega}$ represents the event space, $\mathbb{F}$ denotes
a $\sigma$-algebra of the subsets of ${\bf \Omega}$ and represents
the event space, $\mathbb{F}_{t}$ is a complete filtration given
by the family of $\sigma$-algebras up to time $t$, i.e., $\mathbb{F}_{S}:\mathbb{F}_{S}\subseteq\mathbb{F}_{t}\,\forall t\in\left[0,t\right]$,
and ${\rm P}$ is a probability measure, where the filtration is complete
in the sense it includes all events with probability measure zero
(see \cite{Lanchares.Haddad2023}). Consider a probability space of
$\left({\bf \Omega},\,\mathbb{F},\,{\rm P}\right)$. Then, for any
events $A,\,B\in\mathbb{F}$ such that $A\subseteq B$, the monotonicity
property states that \cite[eq. 2.5]{Billingsley2017} 
\begin{equation}
{\rm P}\left(A\right)\leq{\rm P}\left(B\right).\label{eq:monotonicity}
\end{equation}
 Consider the dynamical system in (\ref{eq: general SDE}). Then,
for some function $V\in\mathcal{C}^{2}$ associated with the process
in (\ref{eq: general SDE}), let the infinitesimal generator $\mathcal{L}$
of the function $V\left(x\right)$ be defined as \cite[eq. 4.12]{Kushner1967}
\begin{gather}
\mathcal{L}V\tq\frac{\partial V}{\partial x}f\left(x\right)+\frac{1}{2}\text{tr}\left(G\left(x,t\right)^{\top}\frac{\partial^{2}V}{\partial x^{2}}G\left(x,t\right)\right).\label{eq: LV_L-1}
\end{gather}

\subsection{Deep Neural Network Model}

Let $\kappa\in\mathbb{R}^{L_{0}}$ denote the DNN input, and $\theta\in\mathbb{R}^{p}$
denote the vector of DNN parameters (i.e., weights and bias terms).
A fully-connected feedforward DNN $\Phi(\kappa,\theta)$ with $k\in\mathbb{Z}_{>0}$
hidden layers and output size $L_{k+1}\in\mathbb{Z}_{>0}$ is defined
using a recursive relation $\varphi_{j}\in\mathbb{R}^{L_{j+1}}$ modeled
as
\begin{eqnarray}
\varphi_{j} & \triangleq & \begin{cases}
V_{j+1}^{\top}\kappa_{a}, & j=0,\\
V_{j+1}^{\top}\phi_{j}\left(\varphi_{j-1}\right) & j\in\left\{ 1,\ldots,k\right\} ,
\end{cases}\label{eq:DNN}
\end{eqnarray}
where $\Phi(\kappa,\theta)=\varphi_{k}$ , $\kappa_{a}\tq\left[\kappa^{\top},1\right]^{\top}$
denotes the augmented input that accounts or the bias terms, $L_{j}\in\mathbb{Z}_{>0}$
denotes the number of neurons in the $j^{\textrm{th}}$ layer with
$L_{j}^{a}\tq L_{j}+1$, and $V_{j+1}\in\mathbb{R}^{L_{j}^{a}\times L_{j+1}}$
denotes the matrix of weights and biases, for all $j\in\left\{ 0,\ldots,k\right\} $.

The vector of activation functions is denoted by $\phi_{j}:\mathbb{R}^{L_{j}}\to\mathbb{R}^{L_{j}^{a}}$
for all $j\in\left\{ 1,\ldots,k\right\} $. The vector of activation
functions can be composed of various activation functions, and hence,
may be represented as $\phi_{j}=\left[\varsigma_{1},\ldots,\varsigma_{L_{j}},1\right]^{\top}$
for all $j\in\left\{ 1,\ldots,k\right\} $, where $\varsigma_{j}:\mathbb{R}\to\mathbb{R}$
for all $j\in\left\{ 1,\ldots,L_{j}\right\} $ denotes a piece-wise
continuously differentiable activation function, where 1 denotes the
augmented hidden layer that accounts for the bias terms. For the DNN
architecture in (\ref{eq:DNN}), the vector of DNN weights is $\theta\triangleq\left[\mathrm{vec}\left(V_{1}\right){}^{\top},\ldots,\mathrm{vec}\left(V_{k}\right){}^{\top}\right]^{\top}$
with size $p=\sum_{j=0}^{k}L_{j}^{a}L_{j+1}$.

Consider $y_{j}\in\mathbb{R}^{L_{j}}$ where $y_{j}=\left[y_{1},\ldots,y_{L_{j}}\right]$
with $y_{i}\in\mathbb{R}$ for all $i\in\left\{ 1,\ldots,L_{j}\right\} $.
The Jacobian $\frac{\partial\phi_{j}}{\partial y_{j}}:\mathbb{R}^{L_{j}}\to\mathbb{R}^{L_{j}^{a}\times L_{j}}$
of the activation function vector at the $j^{\mathrm{th}}$ layer
is given by $\left[\varsigma_{1}^{\prime}(y_{1})\eta_{1},\ldots,\varsigma_{L_{j}}^{\prime}(y_{L_{j}})\eta_{L_{j}},\mathbf{0}_{L_{j}}\right]^{\top}\in\mathbb{R}^{L_{j}^{a}\times L_{j}}$,
where $\varsigma_{j}^{\prime}$ denotes the derivative of $\varsigma_{j}$
with respect to its argument for $j\in\left\{ 1,\ldots,L_{j}\right\} $,
$\eta_{i}$ is the $i^{\text{th}}$ standard basis vector in $\mathbb{R}^{L_{j}}$,
and $\mathbf{0}_{L_{j}}$ is the zero vector in $\mathbb{R}_{L_{j}}$.

Let the gradient of the DNN with respect to the weights be denoted
by $\Phi^{\prime}(\kappa,\theta)\triangleq\frac{\partial}{\partial\theta}\Phi(\kappa,\theta)$,
which can be represented as $\Phi^{\prime}(\kappa,\theta)=\left[\frac{\partial}{\partial{\rm vec}(V_{1})}\Phi(\kappa,\theta),\ldots,\frac{\partial}{\partial{\rm vec}(V_{k+1})}\Phi(\kappa,\theta)\right]\in\mathbb{R}^{L_{k+1}\times p}$,
where $\frac{\partial}{\partial{\rm vec}\left(V_{j}\right)}\Phi\left(\kappa,\theta\right)\in\mathbb{R}^{L_{k+1}\times L_{j-1}^{a}L_{j}}$
for all $j\in\left\{ 1,\ldots,k+1\right\} $. Then, using (\ref{eq:DNN})
and the property of the vectorization operator in (\ref{eq:vec_diff_prop})
yields
\begin{align}
\Phi^{\prime}(\kappa,\theta) & =\left(\stackrel{\curvearrowleft}{\stackrel[\ell=j+1]{k}{\prod}}V_{\ell+1}^{\top}\frac{\partial\phi_{\ell}}{\partial\varphi_{\ell-1}}\right)\left(I_{L_{j+1}}\otimes\varrho_{j}\right),\label{eq:DNN Gradient}
\end{align}
for $j\in\left\{ 0,\ldots,k\right\} $, where $\varrho_{j}=\kappa_{a}^{\top}$
if $j=0$ and $\varrho_{j}=\phi_{j}^{\top}\left(\varphi_{j-1}\right)$
if $j\in\left\{ 1,\ldots,k\right\} $.

\section{Dynamics and Control Objective\label{sec: Dynamics-and-Control}}

Consider a stochastic process modeled by the control-affine nonlinear
stochastic differential equation
\begin{equation}
{\rm d}x=\left(f\left(x\right)+g_{1}\left(x\right)u\left(t\right)\right){\rm d}t+g_{2}\left(x\right)\Sigma\left(t\right){\rm d}\omega,\label{eq: dynamics}
\end{equation}
where $t\in\RR_{\geq0}$ denotes time, $x\in\RR^{n}$ denotes the
known state variable, $f:\RR^{n}\to\RR^{n}$ denotes an unknown continuous
drift vector field, $g_{1}:\RR^{n}\to\RR^{n\times r}$ denotes the
known control effectiveness matrix, and $u\in\RR^{r}$ denotes the
control input. Additionally, in (\ref{eq: dynamics}), $g_{2}:\RR^{n}\to\RR^{n\times s}$
denotes the continuous diffusion matrix, $\Sigma:\RR_{\geq0}\to\RR^{s\times s}$
denotes the symmetric Borel measurable covariance matrix, and $\omega\in\RR^{s}$
denotes the $s$-dimensional independent standard Wiener process defined
on the complete filtered probability space $\left({\bf \Omega},\,\mathbb{F},\,\mathbb{F}_{t},{\rm P}\right)$,
respectively. 
\begin{assumption}
\label{thm: g1} The control effectiveness matrix, $g_{1}$, is full
row rank, and bounded.
\end{assumption}
The objective of this paper is to design a controller such that the
state $x$ converges (in expectation) to a $\mathcal{C}^{2}$-smooth
user-defined desired trajectory $x_{d}:\RR_{\geq0}\to\RR^{n}$. To
quantify the control objective, the tracking error $e\in\RR^{n}$
is defined as
\begin{equation}
e\tq x-x_{d}.\label{eq: tracking error}
\end{equation}

\begin{assumption}
\label{thm: bounded xd}There exist known constants $\overline{x_{d}},\,\overline{\dot{x}_{d}}\in\RR_{>0}$
such that $\left\Vert x_{d}\right\Vert \leq\overline{x_{d}}$ and
$\left\Vert \dot{x}_{d}\right\Vert \leq\overline{\dot{x}_{d}}$.
\end{assumption}
To adapt to the uncertainties caused by the diffusion matrix in the
subsequent stability analysis, Taylor's theorem is applied to the
vectorized diffusion matrix $g_{2}$, yielding
\begin{equation}
\tvec\left(g_{2}\left(x\right)\right)=\Psi\left(e,x_{d}\right)e+\tvec\left(g_{2}\left(x_{d}\right)\right),\label{eq: g2 model}
\end{equation}
where $\Psi:\RR^{n}\times\RR^{n}\to\RR^{nm\times n}$ is a $\mathcal{C}^{\infty}$-smooth
function, and $\tvec\left(g_{2}\left(x_{d}\right)\right)$ is upper-bounded
as $\left\Vert \tvec\left(g_{2}\left(x_{d}\right)\right)\right\Vert \leq\overline{g}$,
where $\bar{g}\in\mathbb{R}_{>0}$ is unknown.

\section{Control Design\label{sec: Control-Design}}

As previously discussed, the multi-DNN architecture in the subsequent
control development allows for targeted adjustments to individual
uncertain components. Let $\mathbf{x}\triangleq\left[x^{\top},\,x_{d}^{\top}\right]^{\top}\in\mathbb{R}^{2n}$.
The terms $\mathcal{F}_{1}:\RR^{n}\to\RR$, and $\mathcal{F}_{2}:\RR^{2n}\to\RR^{n}$
are introduced based on the subsequent analysis and are defined as
$\mathcal{F}_{1}\left(x\right)\tq\left\Vert \Sigma\right\Vert _{F\infty}^{2}\text{tr}\left(\Psi\left(x\right)^{\top}\Psi\left(x\right)\right)$
and $\mathcal{F}_{2}\left(\mathbf{x}\right)\tq\left\Vert \Sigma\right\Vert _{F\infty}^{2}\Psi\left(x\right)^{\top}\tvec\left(g_{2}\left(x_{d}\right)\right)$,
respectively. Leveraging the universal function approximation properties
offered by DNNs, three separate Lb-DNNs are developed to approximate
$f$, $\mathcal{F}_{1}$, and $\mathcal{F}_{2}$.

\subsection{Deep Neural Network Architecture\label{subsec: Deep-Neural-Networks}}

Define the compact domains
\begin{align}
\Omega_{1} & \tq\left\{ \zeta\in\RR^{n}:\left\Vert \zeta\right\Vert \leq\chi+\overline{x_{d}}\right\} ,\nonumber \\
\Omega_{2} & \triangleq\left\{ \xi\in\RR^{2n}:\left\Vert \xi\right\Vert \leq\chi+2\overline{x_{d}}\right\} ,\label{eq:Omega}
\end{align}
where $\chi\in\mathbb{R}_{>0}$ denotes a bounding constant. Prescribe
$\overline{\varepsilon}_{\ell}\in\RR$, and note that $f\in C\left(\Omega_{1},\mathbb{R}^{n}\right)$,
$\mathcal{F}_{1}\in C\left(\Omega_{1},\mathbb{R}\right)$, and $\mathcal{F}_{2}\in C\left(\Omega_{2},\mathbb{R}^{2n}\right)$,
for all $t\in\RR_{\geq0}$ and $\ell\in\left\{ 1,2,3\right\} $. By
\cite[Thm. 3.2]{Kidger.Lyons2020}, there exist Lb-DNNs such that
$\sup_{x\in\Omega_{1}}\left\Vert \Phi_{1}\left(x,\theta_{1}^{*}\right)-f\left(x\right)\right\Vert \leq\overline{\varepsilon}_{1}$,
$\sup_{x\in\Omega_{1}}\left\Vert \Phi_{2}\left(x,\theta_{2}^{*}\right)-\mathcal{F}_{1}\left(x\right)\right\Vert \leq\overline{\varepsilon}_{2}$,
and $\sup_{\mathbf{x}\in\Omega_{2}}\left\Vert \Phi_{3}\left(\mathbf{x},\theta_{3}^{*}\right)-\mathcal{F}_{2}\left(\mathbf{x}\right)\right\Vert \leq\overline{\varepsilon}_{3}$.
Therefore, the uncertainties can be modeled using three separate Lb-DNNs
as 
\begin{align}
f\left(x\right) & =\Phi_{1}\left(x,\theta_{1}^{*}\right)+\varepsilon_{1}\left(x\right),\label{eq: DNN model of f}\\
\mathcal{F}_{1}\left(x\right) & =\Phi_{2}\left(x,\theta_{2}^{*}\right)+\varepsilon_{2}\left(x\right),\label{eq: DNN model of F1}\\
\mathcal{F}_{2}\left(\mathbf{x}\right) & =\Phi_{3}\left(\mathbf{x},\theta_{3}^{*}\right)+\varepsilon_{3}\left(\mathbf{x}\right),\label{eq: DNN model of F2}
\end{align}
for all $x\in\Omega_{1}$ and $\mathbf{x}\in\Omega_{2}$, where $\Phi_{1}:\mathbb{R}^{n}\times\RR^{p_{1}}\to\RR^{n}$,
$\Phi_{2}:\mathbb{R}^{n}\times\RR^{p_{2}}\to\RR$, $\Phi_{3}:\RR^{2n}\times\RR^{p_{3}}\to\RR^{n}$,
and $\varepsilon_{1}:\RR^{n}\to\RR^{n}$, $\varepsilon_{2}:\RR^{n}\to\RR$,
and $\varepsilon_{3}:\RR^{2n}\to\RR^{n}$ denote the unknown function
reconstruction errors that can be bounded as $\sup_{x\in\Omega_{1}}\left\Vert \varepsilon_{1}\left(x\right)\right\Vert \leq\overline{\varepsilon}_{1}$,
$\sup_{x\in\Omega_{1}}\left\Vert \varepsilon_{2}\left(x\right)\right\Vert \leq\overline{\varepsilon}_{2}$,
and $\sup_{{\bf x}\in\Omega_{2}}\left\Vert \varepsilon_{3}\left({\bf x}\right)\right\Vert \leq\overline{\varepsilon}_{3}$.
The following standard assumption is made to aid in the subsequent
development (cf., \cite[Assumption 1]{Lewis.Yesildirek.ea1996}).
\begin{assumption}
\label{thetaBound}For all $\ell\in\left\{ 1,2,3\right\} $, the vector
of ideal weights can be bounded as $\left\Vert \theta_{\ell}^{*}\right\Vert \leq\overline{\theta}_{\ell}$,
where $\overline{\theta}_{\ell}\in\RR_{>0}$ is a known bound. For
cases where $\bar{\theta}_{\ell}$ is unknown, results such as \cite{Fan2018}
can be used to compensate for $\bar{\theta}_{\ell}$.
\end{assumption}

\subsection{Adaptive Update Law \label{subsec: Adaptive-Update-Law}}

Let $\hat{\theta}_{\ell}$ denote the adaptive estimates of the ideal
weights $\theta_{\ell}^{*}$ for all $\ell\in\left\{ 1,2,3\right\} $,
and let the corresponding weight estimation errors be defined as
\begin{equation}
\tilde{\theta}_{\ell}\tq\theta_{\ell}^{*}-\hat{\theta}_{\ell},\,\forall\ell\in\left\{ 1,2,3\right\} .\label{eq: estimation error}
\end{equation}
Motivated by the subsequent Lyapunov-based stability analysis, the
weight estimates are updated according to
\begin{align}
\dot{\hat{\theta}}_{1} & \tq\text{proj}\left(\gamma_{1}\Phi_{1}^{\prime\top}\left(x,\hat{\theta}_{1}\right)e-\gamma_{1}\sigma_{1}\hat{\theta}_{1}\right),\label{eq: update law f}\\
\dot{\hat{\theta}}_{2} & \tq\text{proj}\left(\frac{1}{2}\gamma_{2}e^{\top}e\Phi_{2}^{\prime\top}\left(x,\hat{\theta}_{2}\right)-\gamma_{2}\sigma_{2}\hat{\theta}_{2}\right),\label{eq: update law F1}\\
\dot{\hat{\theta}}_{3} & \tq\text{proj}\left(\gamma_{3}\Phi_{3}^{\prime\top}\left(\mathbf{x},\hat{\theta}_{3}\right)e-\gamma_{3}\sigma_{3}\hat{\theta}_{3}\right),\label{eq: update law F2}
\end{align}
where $\gamma_{\ell}\in\mathbb{R}_{>0}$ denote user-defined learning
rates, $\sigma_{\ell}\in\mathbb{R}_{>0}$ denote user-defined forgetting
factors, and $\text{proj}\left(\cdot\right)$ denotes the smooth projection
operator defined in \cite[eq. (7)-(11)]{Cai2006a}, which is used
to ensure that $\hat{\theta}_{\ell}$ is bounded as $\hat{\theta}_{\ell}\leq\overline{\theta}_{\ell}$,
for all $\ell\in\left\{ 1,2,3\right\} $. 

To facilitate the subsequent stability analysis, a first-order Taylor
approximation \cite[Eq. 22]{Lewis.Yesildirek.ea1996,Patil.Le.ea.2022}
is used on $\Phi_{\ell}-\widehat{\Phi}_{l}$ to yield
\begin{gather}
\Phi_{\ell}-\widehat{\Phi}_{l}=\widehat{\Phi}_{l}^{\prime}\tilde{\theta}_{\ell}+\mathcal{O}_{l}\left(\left\Vert \tilde{\theta}_{\ell}\right\Vert ^{2}\right),\quad\forall\ell\in\left\{ 1,2,3\right\} .\label{eq: TSA-1}
\end{gather}
By Assumption \ref{thetaBound}, boundedness of $\hat{\theta}_{\ell}$,
and given bounded $x$, there exist constants $\Delta_{\ell}\in\mathbb{R}_{>0}$
such that {\large{}$\left\Vert \mathcal{O}_{l}\left(\left\Vert \tilde{\theta}_{\ell}\right\Vert ^{2}\right)\right\Vert \leq\Delta_{l}$},
for all $\ell\in\left\{ 1,2,3\right\} $.

\subsection{Controller and Closed-Loop Error System}

To compensate for the uncertainties that appear in the subsequent
closed-loop error system, the Lb-DNNs are incorporated into the developed
controller, designed as{\scriptsize{}
\begin{equation}
u\left(t\right)\tq g_{1}\left(x\right)^{+}\left(\dot{x}_{d}-k_{e}e-\Phi_{1}\left(x,\hat{\theta}_{1}\right)-\frac{1}{2}e\Phi_{2}\left(x,\hat{\theta}_{2}\right)-\Phi_{3}\left(\mathbf{x},\hat{\theta}_{3}\right)\right),\label{eq: controller}
\end{equation}
}where $k_{e}\in\RR_{>0}$ is a user-defined gain, and $g_{1}^{+}$
exists by Assumption \ref{thm: g1} and \cite{Penrose1955}. Let $z:\RR_{\geq0}\to\RR^{\varphi}$
denote the concatenated error state defined as $z\tq\left[e^{\top},\,\tilde{\theta}_{1}^{\top},\,\tilde{\theta}_{2}^{\top},\,\tilde{\theta}_{3}^{\top}\right]^{\top},$
where $\varphi\tq n+\stackrel[\ell=1]{3}{\sum}p_{\ell}$. Using (\ref{eq: tracking error}),
(\ref{eq: estimation error}), and the chain rule, ${\rm d}z$ is
obtained as ${\rm d}z=\left[{\rm d}x-\frac{{\rm d}x_{d}}{{\rm d}t}{\rm d}t,\,\frac{{\rm d}\tilde{\theta}_{1}}{{\rm d}t}{\rm d}t,\,\frac{{\rm d}\tilde{\theta}_{2}}{{\rm d}t}{\rm d}t,\,\frac{{\rm d}\tilde{\theta}_{3}}{{\rm d}t}{\rm d}t\right].$
Substituting (\ref{eq: dynamics}), (\ref{eq: update law f})-(\ref{eq: controller})
into ${\rm d}z$ yields the closed-loop error system
\begin{equation}
{\rm d}z=F\left(z\right){\rm d}t+G\left(z,t\right){\rm d}\omega,\label{eq: SDE}
\end{equation}
where $G\left(z,t\right)\tq\left[g_{2}\left(x\right)\Sigma\left(t\right),\mathbf{0}_{\left(\varphi-n\right)\times m}\right]^{\top}$
and{\footnotesize{}
\[
F\left(z\right)\tq\left[\begin{array}{c}
f\left(x\right)-k_{e}e-\Phi_{1}\left(x,\hat{\theta}_{1}\right)-\frac{1}{2}e\Phi_{2}\left(x,\hat{\theta}_{2}\right)-\Phi_{3}\left(\mathbf{x},\hat{\theta}_{3}\right)\\
-\text{proj}\left(\gamma_{1}\Phi_{1}^{\prime\top}\left(x,\hat{\theta}_{1}\right)e-\gamma_{1}\sigma_{1}\hat{\theta}_{1}\right)\\
-\text{proj}\left(\frac{1}{2}\gamma_{2}e^{\top}e\Phi_{2}^{\prime\top}\left(x,\hat{\theta}_{2}\right)-\gamma_{2}\sigma_{2}\hat{\theta}_{2}\right)\\
-\text{proj}\left(\gamma_{3}\Phi_{3}^{\prime\top}\left(\mathbf{x},\hat{\theta}_{3}\right)e-\gamma_{3}\sigma_{3}\hat{\theta}_{3}\right)
\end{array}\right].
\]
}{\footnotesize\par}

\section{Stability Analysis\label{sec: Stability-Analysis}}

Let $\mathcal{D}\triangleq\left\{ \iota\in\mathbb{R}^{\varphi}:\left\Vert \iota\right\Vert \leq\chi\right\} $,
where $\chi$ is previously introduced in (\ref{eq:Omega}), and consider
the Lyapunov function candidate $V_{L}:\mathcal{D}\to\RR_{\geq0}$
defined as 
\begin{equation}
V_{L}\left(z\right)\tq\frac{1}{2}e^{\top}e+\frac{1}{2}\tilde{\theta}_{1}^{\top}\gamma_{1}^{-1}\tilde{\theta}_{1}+\frac{1}{2}\tilde{\theta}_{2}^{\top}\gamma_{2}^{-1}\tilde{\theta}_{2}+\frac{1}{2}\tilde{\theta}_{3}^{\top}\gamma_{3}^{-1}\tilde{\theta}_{3}.\label{eq: Lyap. func.}
\end{equation}
The Lyapunov function candidate can be bounded as
\begin{equation}
\alpha_{1}\nzz\leq V_{L}\left(z\right)\leq\alpha_{2}\nzz,\label{eq: V_L bounds}
\end{equation}
where $\alpha_{1}\tq\frac{1}{2}\min\left(1,\gamma_{1}^{-1},\gamma_{2}^{-1},\gamma_{3}^{-1}\right)$
and $\alpha_{2}\tq\max\left(1,\gamma_{1}^{-1},\gamma_{2}^{-1},\gamma_{3}^{-1}\right)$.

The following definition is provided to assist with the subsequent
stability analysis.
\begin{defn}
\label{def: LUUB-p} \textit{(Uniformly ultimately bounded in probability
(UUB-p))} The solutions of (\ref{eq: SDE}) are uniformly ultimately
bounded in probability with bound $\mathtt{b}\in\mathbb{R}_{>0}$
and escape risk ${\tt e}\in(0,1)$, if there exists $\mathtt{c}\in\RR_{>0}$,
independent of $t_{0}\geq0$, such that for every $\mathtt{a}\in\left(0,\mathtt{c}\right)$,
there is ${\tt T}={\tt T}\left(\mathtt{a},\mathtt{b}\right)\geq0$,
independent of $t_{0}$, such that
\[
\left\Vert z\left(t_{0}\right)\right\Vert \leq\mathtt{a}\Rightarrow{\rm P}\left(\underset{0\leq t\leq\infty}{\sup}\left\Vert z\left(t\right)\right\Vert \geq{\tt b}\right)\leq{\tt e},\,\forall t\geq t_{0}+\mathtt{T}.
\]

\medskip{}
\end{defn}
Based on the subsequent stability analysis, let the set of stabilizing
initial conditions be defined as 
\begin{equation}
\mathcal{S}\tq\left\{ \iota\in\RR^{\varphi}:\left\Vert \iota\right\Vert \leq\sqrt{\frac{1}{\alpha_{2}}}\sqrt{\frac{\alpha_{1}}{\alpha_{2}}\chi^{2}-\frac{b}{c}}\right\} .\label{eq:Set mathcal S}
\end{equation}
The set to which all trajectories converge be defined as 
\begin{equation}
\mathcal{B}\tq\left\{ \zeta\in\RR^{\varphi}:\left\Vert \zeta\right\Vert \leq\sqrt{\frac{\lambda}{\alpha_{1}}}\right\} ,\label{eq:Set mathcal B}
\end{equation}
where $\lambda\in\left[\frac{b}{c},m\right]$, $b\tq\frac{1}{2}\left(\Delta_{1}+\Delta_{3}\right)^{2}+\frac{1}{2}\left\Vert \Sigma\Sigma^{\top}\right\Vert _{\infty}\overline{g}^{2}+\frac{\sigma_{1}}{2}\overline{\theta}_{1}^{2}+\frac{\sigma_{2}}{2}\overline{\theta}_{2}^{2}+\frac{\sigma_{3}}{2}\overline{\theta}_{3}^{2}$,
$c\tq\frac{1}{\alpha_{1}}\min\left\{ k_{e}-\frac{1}{2}-\frac{1}{2}\left(\Delta_{2}+\overline{\varepsilon}_{2}\right),\sigma_{1},\sigma_{2}\right\} $,
and $\sigma_{\ell}$ and $\Delta_{\ell}$ are defined in Section \ref{subsec: Adaptive-Update-Law}.

The probability of $\left\Vert z\left(t\right)\right\Vert \leq\chi$
(i.e., $z\left(t\right)\in\mathcal{D}$) is equivalent to the probability
that the Lyapunov function candidate remains below the threshold $m\tq\alpha_{1}\chi^{2}$.
Specifically, following (\ref{eq: V_L bounds}), the condition $V_{L}\left(z\right)\leq m$
ensures that $\left\Vert z\left(t\right)\right\Vert \leq\chi$.
\begin{lem}
\label{thm:probability}\textup{For the Ito process ${\tt z}\in\RR^{n}$
and function ${\tt V}$, assume }
\end{lem}
\begin{lyxlist}{00.00.0000}
\item [{(A1)}] \label{(A1)--is}${\tt V}$ is non-negative, ${\tt V}\left(0\right)=0$,
and ${\tt V}\in\mathcal{C}^{2}$ over the open and connected set $Q_{m}\tq\left\{ {\tt z}:{\tt V}\left({\tt z}\right)<m\right\} $,
where $m\in\RR_{>0}$ is a bounding constant,
\item [{(A2)}] \label{(A2)--is}${\tt z}\left(t\right)$ is a continuous
strong Markov process defined until at least some $\tau^{\prime}>\tau_{m}=\inf\left\{ t:{\tt z}\left(t\right)\notin Q_{m}\right\} $
with probability one,\footnote{This assumption guarantees the existence of the process up to $\tau^{\prime}$
with probability one, an essential requirement for the validity of
the analysis.}
\end{lyxlist}
If $\mathcal{L}{\tt V}\left({\tt z}\right)\leq-\kappa_{1}{\tt V}\left({\tt z}\right)+\kappa_{2}$
in $Q_{m}$ for $\kappa_{1},\kappa_{2}>0$, then for $\lambda\leq m$,
${\tt z}\left(t\right)$ is UUB-p with the probability{\small{}
\begin{gather*}
{\rm P}\left(\underset{t\leq s<\infty}{\sup}{\tt V}\left({\tt z}\left(s\right)\right)\geq\lambda\right)\leq\frac{1}{m}{\tt V}\left({\tt z}\left(0\right)\right)\\
+\frac{1}{\lambda}{\tt V}\left({\tt z}\left(0\right)\right)\exp\left(-\kappa_{1}t\right)+\frac{\kappa_{2}}{\kappa_{1}\lambda}.
\end{gather*}
}{\small\par}
\begin{IEEEproof}
See Appendix.
\end{IEEEproof}
To facilitate the subsequent analysis, the following gain condition
is introduced
\begin{equation}
k_{e}>\frac{1}{2}+\frac{1}{2}\left(\Delta_{2}+\overline{\varepsilon}_{2}\right).\label{eq: gain condition}
\end{equation}
Additionally, let the escape risk of $z$ be defined as
\begin{equation}
\vartheta\tq\frac{1}{m}V_{L}\left(z\left(0\right)\right)+\frac{1}{\lambda}V_{L}\left(z\left(0\right)\right)e^{-ct}+\frac{b}{c\lambda}.\label{eq: vartheta}
\end{equation}
To ensure $\vartheta\in\left(0,1\right)$ and the set $\mathcal{S}$
defined in (\ref{eq:Set mathcal S}) is non-empty, the following feasibility
condition is introduced
\begin{equation}
\chi\geq\sqrt{\frac{\alpha_{2}}{\alpha_{1}}\frac{b}{c}}\sqrt{\frac{\alpha_{2}}{\alpha_{1}}+1}.\label{eq: feasibility condition}
\end{equation}
Recall the set of stabilizing initial conditions in (\ref{eq:Set mathcal S}).
In the subsequent analysis, it is shown that if $z\left(0\right)\in\mathcal{S}\subseteq\mathcal{D}$,
then $z\left(t\right)$ is UUB-p and does not escape $\mathcal{D}$
with a probability with the bound of $1-\vartheta$. The following
theorem states the main result of this paper.
\begin{thm}
\label{thm: stability }Consider the stochastic dynamical system in
(\ref{eq: dynamics}). Let (\ref{eq: gain condition}) and (\ref{eq: feasibility condition})
hold. For any initial conditions of the states $z\left(0\right)\in\mathcal{S}$,
the update laws and controller given by (\ref{eq: update law f})-(\ref{eq: controller})
ensure that the solution $z\left(t\right)$ is UUB-p in the sense
that{\footnotesize{}
\begin{equation}
{\rm P}\left(\underset{t\leq s<\infty}{\sup}\left\Vert z\left(s\right)\right\Vert <\sqrt{\frac{\lambda}{\alpha_{1}}}\right)\geq1-\vartheta.\label{eq:solution}
\end{equation}
}{\footnotesize\par}
\end{thm}
\begin{IEEEproof}
Taking the infinitesimal generator of the candidate Lyapunov function
in (\ref{eq: Lyap. func.}) {\small{}yields} 
\begin{gather}
\mathcal{L}V_{L}\left(z\right)=\frac{\partial V_{L}}{\partial z}F\left(z\right)+\frac{1}{2}\text{tr}\left(G\left(z,t\right)^{\top}\frac{\partial^{2}V_{L}}{\partial z^{2}}G\left(z,t\right)\right).\label{eq: LV_L}
\end{gather}
Substituting $F\left(z\right)$, $G\left(z,t\right)$, and values
of $\frac{\partial V_{L}}{\partial e}$ and $\frac{\partial^{2}V_{L}}{\partial e^{2}}$
into (\ref{eq: LV_L}), using the fact that $e^{\top}e$ is a scalar,
applying the trace property in (\ref{eq: order of multiplication property}),
and incorporating (\ref{eq: estimation error}) and the fact that
$\theta_{\ell}^{*}$ is a constant for all $\ell\in\left\{ 1,2,3\right\} $
yields{\footnotesize{}
\begin{gather}
\mathcal{L}V_{L}=e^{\top}\left(f\left(x\right)-k_{e}e-\Phi_{1}\left(x,\hat{\theta}_{1}\right)-\frac{1}{2}e\Phi_{2}\left(x,\hat{\theta}_{2}\right)-\Phi_{3}\left(\mathbf{x},\hat{\theta}_{3}\right)\right)\nonumber \\
+\frac{1}{2}\text{tr}\left(g_{2}^{\top}g_{2}\Sigma\Sigma^{\top}\right)-\tilde{\theta}_{1}^{\top}\gamma_{1}^{-1}\text{proj}\left(\gamma_{1}\Phi_{1}^{\prime\top}\left(x,\hat{\theta}_{1}\right)e-\gamma_{1}\sigma_{1}\hat{\theta}_{1}\right)\nonumber \\
-\tilde{\theta}_{2}^{\top}\gamma_{2}^{-1}\text{proj}\left(\frac{1}{2}\gamma_{2}e^{\top}e\Phi_{2}^{\prime\top}\left(x,\hat{\theta}_{2}\right)-\gamma_{2}\sigma_{2}\hat{\theta}_{2}\right)\nonumber \\
-\tilde{\theta}_{3}^{\top}\gamma_{3}^{-1}\text{proj}\left(\gamma_{3}\Phi_{3}^{\prime\top}\left(\mathbf{x},\hat{\theta}_{3}\right)e-\gamma_{3}\sigma_{3}\hat{\theta}_{3}\right).\label{eq: LV_L 2}
\end{gather}
}Using the definition of the Frobenius norm on the term $\text{tr}\left(g_{2}^{\top}g_{2}\Sigma\Sigma^{\top}\right)$
yields {\footnotesize{}
\begin{gather}
\text{tr}\left(g_{2}^{\top}g_{2}\Sigma\Sigma^{\top}\right)=\text{tr}\left(\Sigma^{\top}g_{2}^{\top}g_{2}\Sigma\right)=\text{tr}\left(\left(g_{2}\Sigma\right)^{\top}g_{2}\Sigma\right)=\left\Vert g_{2}\Sigma\left(t\right)\right\Vert _{F}^{2}.\label{eq: ggss}
\end{gather}
}Applying the Cauchy-Schwarz inequality \cite[Page 189]{Axler2024}
to (\ref{eq: ggss}) yields
\begin{equation}
\left\Vert g_{2}\Sigma\left(t\right)\right\Vert _{F}^{2}\leq\left\Vert g_{2}\right\Vert _{F}^{2}\left\Vert \Sigma\left(t\right)\right\Vert _{F}^{2}=\text{tr}\left(g_{2}^{\top}g_{2}\right)\left\Vert \Sigma\left(t\right)\right\Vert _{F}^{2}.\label{eq: holder on ggss}
\end{equation}
Using (\ref{eq: separating trace inequality}), (\ref{eq: ggss}),
(\ref{eq: holder on ggss}), and the fact that $\left\Vert \Sigma\left(t\right)\right\Vert _{F}^{2}\leq\underset{t\in\RR_{\geq0}}{\sup}\left\Vert \Sigma\left(t\right)\right\Vert _{F}^{2}\triangleq\left\Vert \Sigma\right\Vert _{F\infty}^{2}$,
the term $\text{tr}\left(g_{2}^{\top}g_{2}\Sigma\Sigma^{\top}\right)$
in (\ref{eq: LV_L 2}) is upper bounded as 
\begin{equation}
\text{tr}\left(g_{2}^{\top}g_{2}\Sigma\Sigma^{\top}\right)\leq\text{tr}\left(g_{2}^{\top}g_{2}\right)\left\Vert \Sigma\right\Vert _{F\infty}^{2}.\label{eq: ggss final}
\end{equation}
Thus, applying (\ref{eq: ggss final}) to (\ref{eq: LV_L 2}), applying
the trace-to-vector property in (\ref{eq: trace to vec}) to $\text{tr}\left(g_{2}^{\top}g_{2}\right)$,
and substituting (\ref{eq: g2 model}) into (\ref{eq: LV_L 2}) yields
{\footnotesize{}
\begin{gather}
\mathcal{L}V_{L}=e^{\top}\left(f\left(x\right)-k_{e}e-\Phi_{1}\left(x,\hat{\theta}_{1}\right)-\frac{1}{2}e\Phi_{2}\left(x,\hat{\theta}_{2}\right)-\Phi_{3}\left(\mathbf{x},\hat{\theta}_{3}\right)\right)\nonumber \\
+\frac{1}{2}\left\Vert \Sigma\right\Vert _{F\infty}^{2}\bigg(e^{\top}\Psi\left(x\right)^{\top}\Psi\left(x\right)e+2e^{\top}\Psi\left(x\right)^{\top}\tvec\left(g_{2}\left(x_{d}\right)\right)\nonumber \\
+\tvec\left(g_{2}\left(x_{d}\right)\right)^{\top}\tvec\left(g_{2}\left(x_{d}\right)\right)\bigg)\nonumber \\
-\tilde{\theta}_{1}^{\top}\gamma_{1}^{-1}\text{proj}\left(\gamma_{1}\Phi_{1}^{\prime\top}\left(x,\hat{\theta}_{1}\right)e-\gamma_{1}\sigma_{1}\hat{\theta}_{1}\right)\nonumber \\
-\tilde{\theta}_{2}^{\top}\gamma_{2}^{-1}\text{proj}\left(\frac{1}{2}\gamma_{2}e^{\top}e\Phi_{2}^{\prime\top}\left(x,\hat{\theta}_{2}\right)-\gamma_{2}\sigma_{2}\hat{\theta}_{2}\right)\nonumber \\
-\tilde{\theta}_{3}^{\top}\gamma_{3}^{-1}\text{proj}\left(\gamma_{3}\Phi_{3}^{\prime\top}\left(\mathbf{x},\hat{\theta}_{3}\right)e-\gamma_{3}\sigma_{3}\hat{\theta}_{3}\right).\label{eq: LV_L 3}
\end{gather}
}Applying the Cauchy-Schwarz inequality to $\frac{1}{2}\left\Vert \Sigma\right\Vert _{F\infty}^{2}e^{\top}\Psi^{\top}\Psi e$
yields
\begin{gather}
\frac{1}{2}\left\Vert \Sigma\right\Vert _{F\infty}^{2}e^{\top}\Psi^{\top}\Psi e=\frac{1}{2}\left\Vert \Sigma\right\Vert _{F\infty}^{2}\left\Vert \Psi e\right\Vert ^{2}\nonumber \\
\leq\frac{1}{2}\left\Vert \Sigma\right\Vert _{F\infty}^{2}\left\Vert \Psi\right\Vert _{F}^{2}\left\Vert e\right\Vert ^{2}=\frac{1}{2}\left\Vert \Sigma\right\Vert _{F\infty}^{2}e^{\top}e\text{tr}\left\{ \Psi\left(x\right)^{\top}\Psi\left(x\right)\right\} .\label{eq: psipsi}
\end{gather}
Thus, applying (\ref{eq: psipsi}) and (\ref{eq: DNN model of f})-(\ref{eq: DNN model of F2})
to (\ref{eq: LV_L 3}), and upper-bounding $\tvec\left(g_{2}\left(x_{d}\right)\right)$
as $\tvec\left(g_{2}\left(x_{d}\right)\right)\leq\left\Vert \tvec\left(g_{2}\left(x_{d}\right)\right)\right\Vert \leq\overline{g}$,
yields{\scriptsize{}
\begin{gather}
\mathcal{L}V_{L}=e^{\top}\bigg(\Phi_{1}\left(x,\theta_{1}\right)+\varepsilon_{1}\left(x\right)-k_{e}e-\Phi_{1}\left(x,\hat{\theta}_{1}\right)-\frac{1}{2}e\Phi_{2}\left(x,\hat{\theta}_{2}\right)\nonumber \\
-\Phi_{3}\left(\mathbf{x},\hat{\theta}_{3}\right)\bigg)+\frac{1}{2}e^{\top}e\left(\Phi_{2}\left(x,\theta_{2}\right)+\varepsilon_{2}\left(x\right)\right)+e^{\top}\left(\Phi_{3}\left(\mathbf{x},\theta_{3}\right)+\varepsilon_{3}\left(\mathbf{x}\right)\right)\nonumber \\
+\frac{1}{2}\left\Vert \Sigma\right\Vert _{F\infty}^{2}\overline{g}^{2}-\tilde{\theta}_{1}^{\top}\gamma_{1}^{-1}\text{proj}\left(\gamma_{1}\Phi_{1}^{\prime\top}\left(x,\hat{\theta}_{1}\right)e-\gamma_{1}\sigma_{1}\hat{\theta}_{1}\right)\nonumber \\
-\tilde{\theta}_{2}^{\top}\gamma_{2}^{-1}\text{proj}\left(\frac{1}{2}\gamma_{2}e^{\top}e\Phi_{2}^{\prime\top}\left(x,\hat{\theta}_{2}\right)-\gamma_{2}\sigma_{2}\hat{\theta}_{2}\right)\nonumber \\
-\tilde{\theta}_{3}^{\top}\gamma_{3}^{-1}\text{proj}\left(\gamma_{3}\Phi_{3}^{\prime\top}\left(\mathbf{x},\hat{\theta}_{3}\right)e-\gamma_{3}\sigma_{3}\hat{\theta}_{3}\right).\label{eq: LV_L 4-1}
\end{gather}
}From \cite[P2 in Thm. 1]{Cai2006a}, $\gamma_{3}\Phi_{3}^{\prime\top}\left(\mathbf{x},\hat{\theta}_{3}\right)e-\gamma_{3}\sigma_{3}\hat{\theta}_{3}\leq\text{proj}\left(\gamma_{3}\Phi_{3}^{\prime\top}\left(\mathbf{x},\hat{\theta}_{3}\right)e-\gamma_{3}\sigma_{3}\hat{\theta}_{3}\right)$.
Applying this inequality to (\ref{eq: LV_L 4-1}), substituting (\ref{eq: TSA-1})
into (\ref{eq: LV_L 4-1}), and cancelling the cross terms, incorporating
(\ref{eq: estimation error}), and utilizing the bounds on the reconstruction
errors and the higher order terms yields
\begin{gather}
\mathcal{L}V_{L}\leq-k_{e}e^{\top}e-\sigma_{1}\tilde{\theta}_{1}^{\top}\tilde{\theta}_{1}-\sigma_{2}\tilde{\theta}_{2}^{\top}\tilde{\theta}_{2}-\sigma_{3}\tilde{\theta}_{3}^{\top}\tilde{\theta}_{3}\nonumber \\
+\left|e^{\top}\right|\left(\Delta_{1}+\overline{\varepsilon}_{1}\right)+\frac{1}{2}e^{\top}e\left(\Delta_{2}+\overline{\varepsilon}_{2}\right)+\left|e^{\top}\right|\left(\Delta_{3}+\overline{\varepsilon}_{3}\right)\nonumber \\
+\frac{1}{2}\left\Vert \Sigma\right\Vert _{F\infty}^{2}\overline{g}^{2}+\sigma_{1}\tilde{\theta}_{1}^{\top}\theta_{1}^{*}+\sigma_{2}\tilde{\theta}_{2}^{\top}\theta_{2}^{*}+\sigma_{3}\tilde{\theta}_{3}^{\top}\theta_{3}^{*},\label{eq: LV_L 4}
\end{gather}
when $z\in\mathcal{D}$. Applying Young's inequality on the terms
$\left|e^{\top}\right|\left(\Delta_{1}+\overline{\varepsilon}_{1}+\Delta_{3}+\overline{\varepsilon}_{3}\right)$,
$\sigma_{\ell}\tilde{\theta}_{\ell}^{\top}\theta_{\ell}^{*}$, for
all $\ell\in\left\{ 1,2,3\right\} $, incorporating $z$ and the gain
condition $k_{e}>\frac{1}{2}+\frac{1}{2}\left(\Delta_{2}+\overline{\varepsilon}_{2}\right)$
leads to 
\[
\mathcal{L}V_{L}\leq-c_{1}\nzz+b,
\]
when $z\in\mathcal{D}$. Using the upper-bound $V_{L}\leq\alpha_{2}\nzz$
introduced in (\ref{eq: V_L bounds}) yields
\begin{equation}
\mathcal{L}V_{L}\leq-cV_{L}+b,\label{eq: LV_L final}
\end{equation}
when $z\in\mathcal{D}$. Since $V_{L}\left(0\right)=0$, $V_{L}\in\mathcal{C}^{2}$,
and $z$ is a continuous strong Markov process, then assumptions (A1)
and (A2) in Lemma \ref{thm:probability} is satisfied. Therefore,
from (\ref{eq: LV_L final}) and Lemma \ref{thm:probability}, ${\rm P}\left(\underset{t\leq s<\infty}{\sup}V_{L}\left(z\left(s\right)\right)\geq\lambda\right)\leq\vartheta$,
when $z\in\mathcal{D}$, which is equivalent to
\begin{equation}
{\rm P}\left(\underset{t\leq s<\infty}{\sup}V_{L}\left(z\left(s\right)\right)<\lambda\right)\geq1-\vartheta.\label{eq: bruhIDC}
\end{equation}
From (\ref{eq: V_L bounds}), ${\rm P}\left(\underset{t\leq s<\infty}{\sup}\left\Vert z\left(s\right)\right\Vert ^{2}<\frac{\lambda}{\alpha_{1}}\right)\geq{\rm P}\left(\underset{t\leq s<\infty}{\sup}V_{L}\left(z\left(s\right)\right)<\lambda\right)$.
Therefore, using (\ref{eq: bruhIDC}) and Lemma \ref{thm:probability}
yields (\ref{eq:solution}), when $z\left(t\right)\in\mathcal{D}$.
From (\ref{eq:solution}) and Definition \ref{def: LUUB-p}, the solution
$z\left(t\right)$ is UUB-p with $z\in\mathcal{D}$.

\begin{figure}
\resizebox{\columnwidth}{!}{%
	\begin{tikzpicture} 
	
	\draw[thick] (0,0) circle (1cm); 
	\draw[thick] (0,0) circle (2.5cm); 
	\draw[dashed,thick, brown] (0,0) circle (3.25cm);
	\draw[thick] (0,0) circle (4cm); 
	\draw[thick] (0,0) circle (5cm);
	
	\node at (1.3cm, 0cm) {\(\mathcal{B}\)}; 
	\node at (2.8cm, 0cm) {\(\mathcal{S}\)}; 
	\node at (4.3cm, 0cm) {\(\mathcal{D}\)};
	\node at (6.1cm, 0cm) {\(\{z : V_L < m\}\)}; 
	
	\draw[->, thick, brown] (0cm, 0cm) -- (-3.25cm, 0cm); 
	\node[below, brown] at (-2cm, 0.5cm) {\(\lambda\)};
	
	\draw[fill=black] (0, 0) circle (1pt); 
	\node[below left, black] at (0.5cm, 0.5cm) {origin};

	\draw[fill=blue] (-2cm, 1) circle (2pt); 
	\node[below left, blue] at (-1.5cm, 1) {start};
	
	\draw[->, blue, thick, decorate, decoration={snake, amplitude=0.5mm, segment length=0.8cm}]    (-2cm, 1)    
	.. controls (-1.5cm, 2cm) and (0cm, 2cm) 
	.. (2cm, 2cm) 
	.. controls (2.5cm, 1cm) and (4cm, 2cm) 
	.. (3cm, -1cm) 
	.. controls (3cm, -1cm) and (1cm, -2.5cm) 
	.. (0cm, -2cm) 
	.. controls (-1.8cm, -1.5cm) and (-2cm, -1cm) 
	.. (-1.2cm, 0.5cm) 
	.. controls (0.2cm, 0.5cm) and (0.9cm, 0cm) 
	.. (0.3cm, -0.3cm) 
	.. controls (0.2cm, -0.8cm) and (0cm, -0.1cm) 
	.. (-0.2cm, -0.2cm); 
	
	
	\draw[->, red, thick, decorate, decoration={snake, amplitude=0.5mm, segment length=0.8cm}]   (-2cm, 1) 
	.. controls (0cm, 2cm) and (1.5cm, 3cm) 
	.. (2.5cm, 4cm) 
	.. controls (3.5cm, 5cm) and (4cm, 5.3cm) 
	.. (4.5cm, 5.8cm); 
	
	\draw[<-, dashed, thick] (3.5cm, 1cm) -- (6cm, 1cm);  
	\node[above, black] at (7.6cm, 0.7cm) {probability of \(1 - \vartheta\)}; 
	
	\draw[<-, dashed, thick] (3.8cm, 5cm) -- (6cm, 5cm);  
	\node[above, black] at (7.3cm, 4.7cm) {probability of \( \vartheta \)};
	
	\end{tikzpicture}
}
\begin{centering}
\caption{\textcolor{blue}{\label{fig:sets}}For a UUB-p system, if the states
are initialized within the set $\mathcal{S}$, they remain inside
the set $\mathcal{D}$ with probability $1-\vartheta$ and eventually
exponentially converge to the set $\mathcal{D}$, staying within the
bounded set (blue trajectory). However, there is an escape risk of
$\vartheta$, meaning the trajectories can potentially become unbounded
(red trajectory). Additionally, $\lambda$ is the radius of an arbitrary
level set, whose size corresponds to either the minimum size of $\mathcal{B}$
or the maximum size of $\left\{ z:V_{L}<m\right\} $.}
\par\end{centering}
\end{figure}
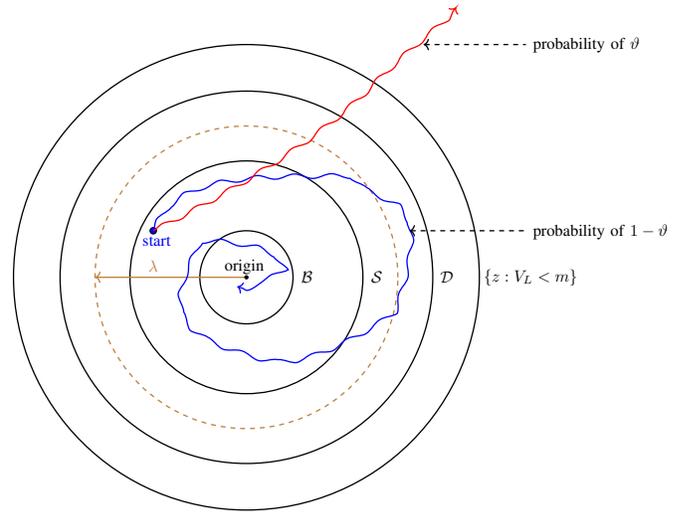

To establish the universal approximation property, it suffices to
show that $\mathbf{x}\in\Omega_{2}$, where $\Omega_{2}$ was introduced
in (\ref{eq:Omega}). By construction, $x_{d}\in\Omega_{1}$. Now,
let $\iota\in\mathbb{R}^{\varphi}$. Since $z\in\mathcal{D},$ $x\in\Omega_{1}$.
Furthermore, since $z\in\mathcal{D}$, $e\in\mathcal{D}$ which implies
$\left\Vert x-x_{d}\right\Vert \leq\chi$. Using triangle inequality,
$\left\Vert x\right\Vert -\left\Vert x_{d}\right\Vert \leq\chi$ which
implies $\left\Vert x\right\Vert \leq\chi+\overline{x_{d}}$. Therefore,
$x\in\Omega_{1}$. Since $\left\{ {\bf x}:x,x_{d}\in\Omega_{1}\right\} \subset\left\{ \mathbf{x}:\mathbf{x}\in\Omega_{2}\right\} $,
it is concluded that $\mathbf{x}\in\Omega_{2}$.

The analysis so far involves the condition $z\left(t\right)\in\mathcal{D}$
(see, Figure \ref{fig:sets}). To obtain conditions with probability
bounds for when $z\left(t\right)\in\mathcal{D}$, let $S_{1}\tq\left\{ z:\left\Vert z\left(t\right)\right\Vert <\sqrt{\frac{\lambda}{\alpha_{1}}}\right\} $
and $S_{2}\tq\left\{ z:\left\Vert z\left(t\right)\right\Vert <\sqrt{\frac{m}{\alpha_{1}}}\right\} $.
Since $S_{1}\subseteq S_{2}$, by substituting $m=\alpha_{1}\chi^{2}$
into $S_{2}$ and invoking the monotonicity property in (\ref{eq:monotonicity}),
${\rm P}\left(\underset{t\leq s<\infty}{\sup}\left\Vert z\left(s\right)\right\Vert <\chi\right)\geq{\rm P}\left(\underset{t\leq s<\infty}{\sup}\left\Vert z\left(s\right)\right\Vert <\sqrt{\frac{\lambda}{\alpha_{1}}}\right)$.
Therefore, this inequality together with (\ref{eq:solution}) yields
${\rm P}\left(\underset{t\leq s<\infty}{\sup}\left\Vert z\left(s\right)\right\Vert <\chi\right)\geq1-\vartheta$
which ensures $z\left(t\right)\in\mathcal{D}$ with the probability
of $1-\vartheta$. 

To ensure the states remain bounded, it needs to be shown that $\mathcal{S}\subseteq\mathcal{D}$.
Consider any $\iota_{1}\in\mathbb{R}^{\varphi}$ and let $\iota_{1}\in\mathcal{S}$.
From (\ref{eq:Set mathcal S}), $\left\Vert \iota_{1}\right\Vert \leq\sqrt{\frac{1}{\alpha_{2}}}\sqrt{\frac{\alpha_{1}}{\alpha_{2}}\chi^{2}-\frac{b}{c}}\leq\sqrt{\frac{1}{\alpha_{2}}}\sqrt{\frac{\alpha_{1}}{\alpha_{2}}}\chi\leq\chi$.
Since $\alpha_{2}>1$ and $\alpha_{1}<\alpha_{2}$, $\left\Vert \iota_{1}\right\Vert \leq\sqrt{\frac{1}{\alpha_{2}}}\sqrt{\frac{\alpha_{1}}{\alpha_{2}}}\chi\leq\chi$,
implying that $\iota_{1}\in\mathcal{D}$. Therefore, $\mathcal{S}\subseteq\mathcal{D}$.

To ensure the states converge to a subset of the set of stabilizing
initial conditions, it suffices to show $\mathcal{B}\subseteq\mathcal{S}$.
Consider any $\iota_{2}\in\mathbb{R}^{\varphi}$ and let $\iota_{2}\in\mathcal{B}$.
Solving the condition $\chi\geq\sqrt{\frac{\alpha_{2}}{\alpha_{1}}\frac{b}{c}}\sqrt{\frac{\alpha_{2}}{\alpha_{1}}+1}$
for $\sqrt{\frac{b}{c\alpha_{1}}}$ yields $\sqrt{\frac{b}{c\alpha_{1}}}\leq\sqrt{\frac{1}{\alpha_{2}}}\sqrt{\frac{\alpha_{1}}{\alpha_{2}}\chi^{2}-\frac{b}{c}}$,
where the obtained inequality together with the definition of $\mathcal{B}$
in (\ref{eq:Set mathcal B}) yields $\left\Vert \iota_{2}\right\Vert \leq\sqrt{\frac{b}{c\alpha_{1}}}\leq\sqrt{\frac{1}{\alpha_{2}}}\sqrt{\frac{\alpha_{1}}{\alpha_{2}}\chi^{2}-\frac{b}{c}}$,
implying that $\iota_{2}\in\mathcal{S}$. Therefore, $\mathcal{B}\subseteq\mathcal{S}$.
Finally, since $\chi\leq\chi+2\overline{x_{d}}$, from the definitions
of sets $\mathcal{D}$ and $\Omega_{2}$, it is ensured that $\mathcal{D}\subseteq\left\{ z:\mathbf{x}\in\Omega_{2}\right\} $.
Therefore, $\mathcal{B}\subseteq\mathcal{S}\subseteq\mathcal{D}\subseteq\left\{ z:\mathbf{x}\in\Omega_{2}\right\} $.

Recall $S_{1}\tq\left\{ z:\left\Vert z\right\Vert <\sqrt{\frac{\lambda}{\alpha_{1}}}\right\} $,
and let $S_{3}\tq\left\{ z:\left\Vert e\right\Vert <\sqrt{\frac{\lambda}{\alpha_{1}}}\right\} $.
Since $S_{1}\subseteq S_{3}$, the monotonicity property in (\ref{eq:monotonicity})
yields ${\rm P}\left(\underset{t\leq s<\infty}{\sup}\left\Vert z\left(s\right)\right\Vert <\sqrt{\frac{\lambda}{\alpha_{1}}}\right)\leq{\rm P}\left(\underset{t\leq s<\infty}{\sup}\left\Vert e\left(s\right)\right\Vert <\sqrt{\frac{\lambda}{\alpha_{1}}}\right).$
This inequality together with (\ref{eq:solution}) yields ${\rm P}\left(\underset{t\leq s<\infty}{\sup}\left\Vert e\left(s\right)\right\Vert <\sqrt{\frac{\lambda}{\alpha_{1}}}\right)\geq1-\vartheta$.
Let $S_{4}\tq\left\{ z:\left\Vert x\right\Vert <\sqrt{\frac{\lambda}{\alpha_{1}}}+\overline{x_{d}}\right\} $.
Since $S_{3}\subseteq S_{4}$, the monotonicity property in (\ref{eq:monotonicity})
yields ${\rm P}\left(\underset{t\leq s<\infty}{\sup}\left\Vert x\left(s\right)\right\Vert <\sqrt{\frac{\lambda}{\alpha_{1}}}+\overline{x_{d}}\right)\geq{\rm P}\left(\underset{t\leq s<\infty}{\sup}\left\Vert e\left(s\right)\right\Vert <\sqrt{\frac{\lambda}{\alpha_{1}}}\right)\geq1-\vartheta$.
Let $S_{5,\ell}\tq\left\{ z:\left\Vert \tilde{\theta}_{\ell}\right\Vert <\sqrt{\frac{\lambda}{\alpha_{1}}}\right\} $,
for all $\ell\in\left\{ 1,2,3\right\} $. Since $S_{1}\subseteq S_{5,\ell}$,
the monotonicity property yields ${\rm P}\left(\underset{t\leq s<\infty}{\sup}\left\Vert \tilde{\theta}_{\ell}\left(s\right)\right\Vert <\sqrt{\frac{\lambda}{\alpha_{1}}}\right)\geq{\rm P}\left(\underset{t\leq s<\infty}{\sup}\left\Vert z\left(s\right)\right\Vert <\sqrt{\frac{\lambda}{\alpha_{1}}}\right)$,
for all $\ell\in\left\{ 1,2,3\right\} $. This obtained inequality
together with (\ref{eq:solution}) yields ${\rm P}\left(\underset{t\leq s<\infty}{\sup}\left\Vert \tilde{\theta}_{\ell}\left(s\right)\right\Vert <\sqrt{\frac{\lambda}{\alpha_{1}}}\right)\geq1-\vartheta$,
for all $\ell\in\left\{ 1,2,3\right\} $. Since ${\rm P}\left\{ \underset{t\leq s<\infty}{\sup}\left\Vert x\left(s\right)\right\Vert <\sqrt{\frac{\lambda}{\alpha_{1}}}+\overline{x_{d}}\right\} \geq1-\vartheta$
and $\hat{\theta}_{\ell}$ is bounded, and based on the smoothness
of $\Phi_{1}\left(x,\hat{\theta}_{1}\right)$, $\Phi_{2}\left(x,\hat{\theta}_{2}\right)$,
and $\Phi_{3}\left(\mathbf{x},\hat{\theta}_{3}\right)$ there exists
a constant $\overline{\Phi_{\ell}}\in\RR_{>0}$ such that ${\rm P}\left(\underset{t\leq s<\infty}{\sup}\left\Vert \Phi_{1}\left(x\left(s\right),\hat{\theta}_{1}\left(s\right)\right)\right\Vert \leq\overline{\Phi_{1}}\right)\geq1-\vartheta$,
${\rm P}\left(\underset{t\leq s<\infty}{\sup}\left\Vert \Phi_{2}\left(x\left(s\right),\hat{\theta}_{2}\left(s\right)\right)\right\Vert \leq\overline{\Phi_{2}}\right)\geq1-\vartheta$,
and ${\rm P}\left(\underset{t\leq s<\infty}{\sup}\left\Vert \Phi_{3}\left(\mathbf{x}\left(s\right),\hat{\theta}_{3}\left(s\right)\right)\right\Vert \leq\overline{\Phi_{3}}\right)\geq1-\vartheta$,
for all $\ell\in\left\{ 1,2,3\right\} $. Since ${\rm P}\left(\underset{t\leq s<\infty}{\sup}\left\Vert \Phi\left(x\left(s\right),\hat{\theta}_{1}\left(s\right)\right)\right\Vert \leq\overline{\Phi_{1}}\right)\geq1-\vartheta$,
${\rm P}\left(\underset{t\leq s<\infty}{\sup}\left\Vert \Phi\left(x\left(s\right),\hat{\theta}_{2}\left(s\right)\right)\right\Vert \leq\overline{\Phi_{2}}\right)\geq1-\vartheta$,
${\rm P}\left(\underset{t\leq s<\infty}{\sup}\left\Vert \Phi\left(\mathbf{x}\left(s\right),\hat{\theta}_{3}\left(s\right)\right)\right\Vert \leq\overline{\Phi_{3}}\right)\geq1-\vartheta$,
and ${\rm P}\left(\underset{t\leq s<\infty}{\sup}\left\Vert e\left(s\right)\right\Vert <\sqrt{\frac{\lambda}{\alpha_{1}}}\right)\geq1-\vartheta$,
using Assumption \ref{thm: bounded xd} and (\ref{eq: controller})
yields ${\rm P}\left(\underset{t\leq s<\infty}{\sup}\left\Vert u\left(s\right)\right\Vert \leq\overline{u}\right)\geq1-\vartheta$,
for some constant $\overline{u}\in\RR_{>0}$. Therefore, all implemented
signals are bounded with probability of $1-\vartheta$.
\end{IEEEproof}

\section{Simulation\label{sec: Simulation}}

To determine the efficacy of the proposed Lb-DNN adaptive controller,
two simulations are performed on a five-dimensional nonlinear stochastic
dynamical system, where $f$, $g_{2}$, and $\Sigma$ in (\ref{eq: dynamics})
are defined as
\begin{align*}
f & =\left[\begin{array}{c}
x_{4}\sqrt{\left|x_{3}\right|}+\sin\left(x_{1}\right)+x_{5}^{2}x_{2}\\
1.5x_{3}^{2}x_{5}+\cos\left(x_{3}+x_{4}\right)+x_{1}\sqrt{\left|x_{2}\right|}\sin\left(x_{3}\right)\\
x_{5}^{2}-x_{3}^{3}x_{4}^{2}\\
\left(x_{1}x_{3}-x_{2}\right)^{3}\\
-x_{1}x_{5}
\end{array}\right],\\
g_{2} & =\left[\begin{array}{cc}
x_{1}\cos\left(x_{2}\right) & 1-x_{3}\cos\left(x_{4}\right)\\
x_{3}x_{5} & x_{4}^{2}\sin^{2}\left(x_{2}\right)\\
x_{1}^{2} & x_{3}\cos\left(x_{1}x_{2}\right)\\
\left(x_{1}+x_{2}\right)^{3}-\sin\left(x_{3}\right) & 1-x_{3}^{2}\\
x_{2}\sin^{2}\left(x_{3}\right) & -x_{5}+x_{1}x_{4}^{2}
\end{array}\right],\\
\Sigma & =\left[\begin{array}{cc}
\sin^{2}\left(t\right) & 0\\
0 & \exp\left(-t\right)
\end{array}\right],
\end{align*}
where $x\tq\left[x_{1},\,x_{2},\,x_{3},\,x_{4},x_{5}\right]^{\top}:\RR_{\geq0}\to\RR^{3}$
denotes the system state, and $g_{1}=I_{5\times5}$. The simulations
are performed for $60$ seconds with initial condition $x\left(0\right)=\left[2,\,-1,\,2,\,-1,\,2\right]^{\top}$.
The desired trajectory is selected as 
\[
x_{d}=\left[\begin{array}{c}
\sin\left(2t\right)\\
-\cos\left(t\right)\\
\sin\left(3t\right)+\cos\left(-2t\right)\\
\sin\left(t\right)-\cos\left(-0.5t\right)\\
\sin\left(-t\right)
\end{array}\right].
\]
Two simulations are performed; the first one represents the performance
of the developed Lb-DNN controller, and the second one illustrates
the tracking perfomance of the controller in response to variations
in mean and covariance of the stochastic noise. For the first simulation,
the Wiener process, $\omega$, is generated with mean of 0 and covariance
of 1, whereas for the second simulation, the noise mean varies from
-0.1 to 0.1 and the noise covariance ranges from 1 to 10. For both
of the simulations, the learning rates and forgetting factors for
each of the Lb-DNNs are selected as $\gamma_{1}=\gamma_{3}=25,\,\gamma_{2}=5$,
$\sigma_{1}=\sigma_{3}=0.01$, and $\sigma_{2}=0.1$, respectively.
For both simulations, the control gain in (\ref{eq: controller})
is selected as $k_{e}=50$0. Both Lb-DNNs have $k_{1}=k_{2}=k_{3}=8$
inner layers with $L_{1}=L_{2}=L_{3}=8$ neurons per hidden layer.
In these simulations, the Lb-DNNs use swish activation functions (see
\cite{Ramachandran.Zoph.ea2017}). Since swish activation, a smooth
approximation of ReLu activation, is used for the simulations, the
weight estimates are initialized via Kaiming He initialization (see
\cite{He.Zhang.ea2015}).

\begin{figure}
\begin{centering}
\includegraphics[viewport=75bp 10bp 1005bp 510bp,clip,scale=0.25]{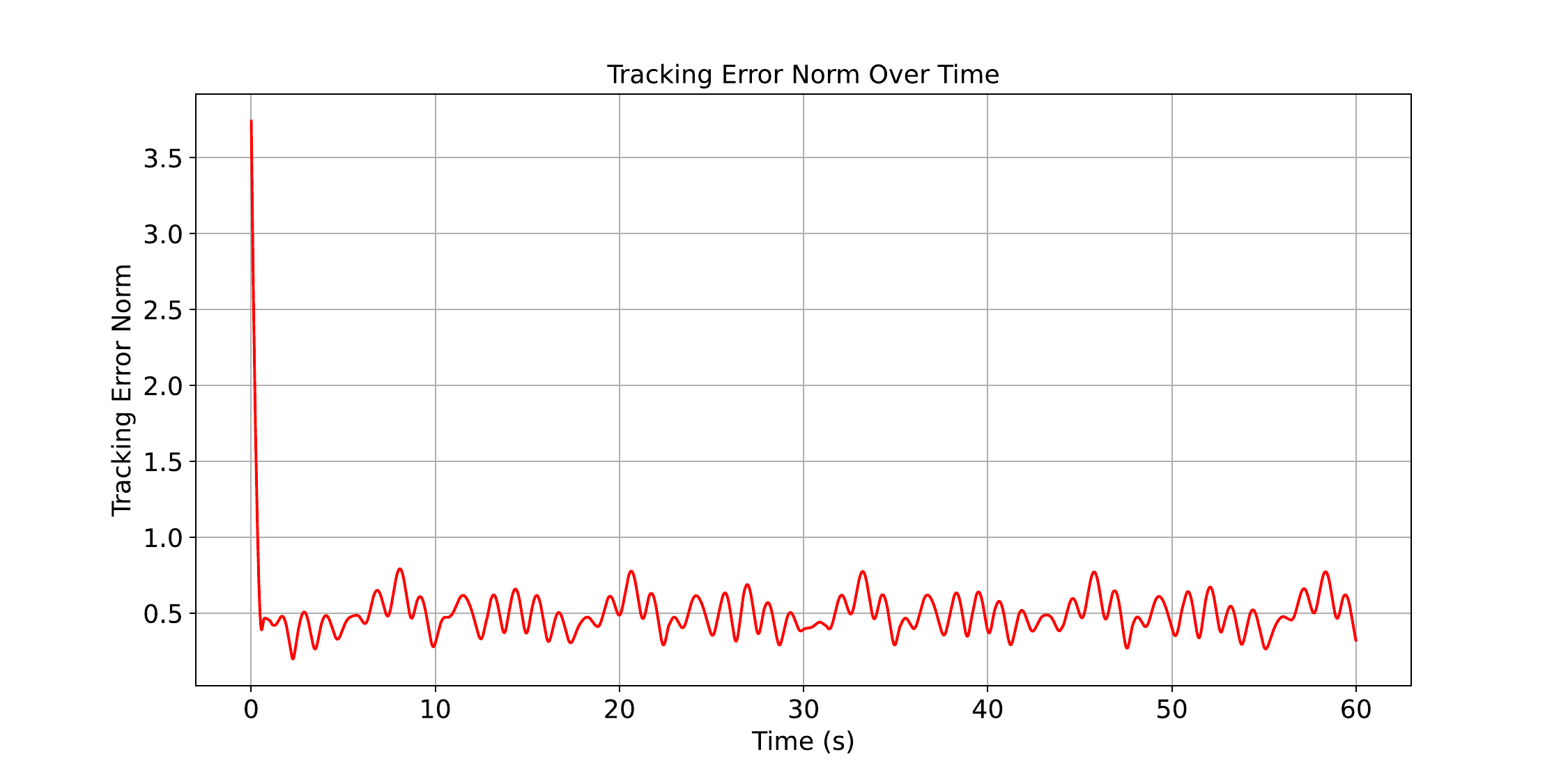}
\par\end{centering}
\centering{}\caption{\label{fig:track. err.}Performance of the tracking error over time
for the developed Lb-DNN controller.}
\end{figure}

\begin{figure}
\begin{centering}
\includegraphics[viewport=4.36237bp 2.180943bp 807.038bp 610.664bp,clip,scale=0.3]{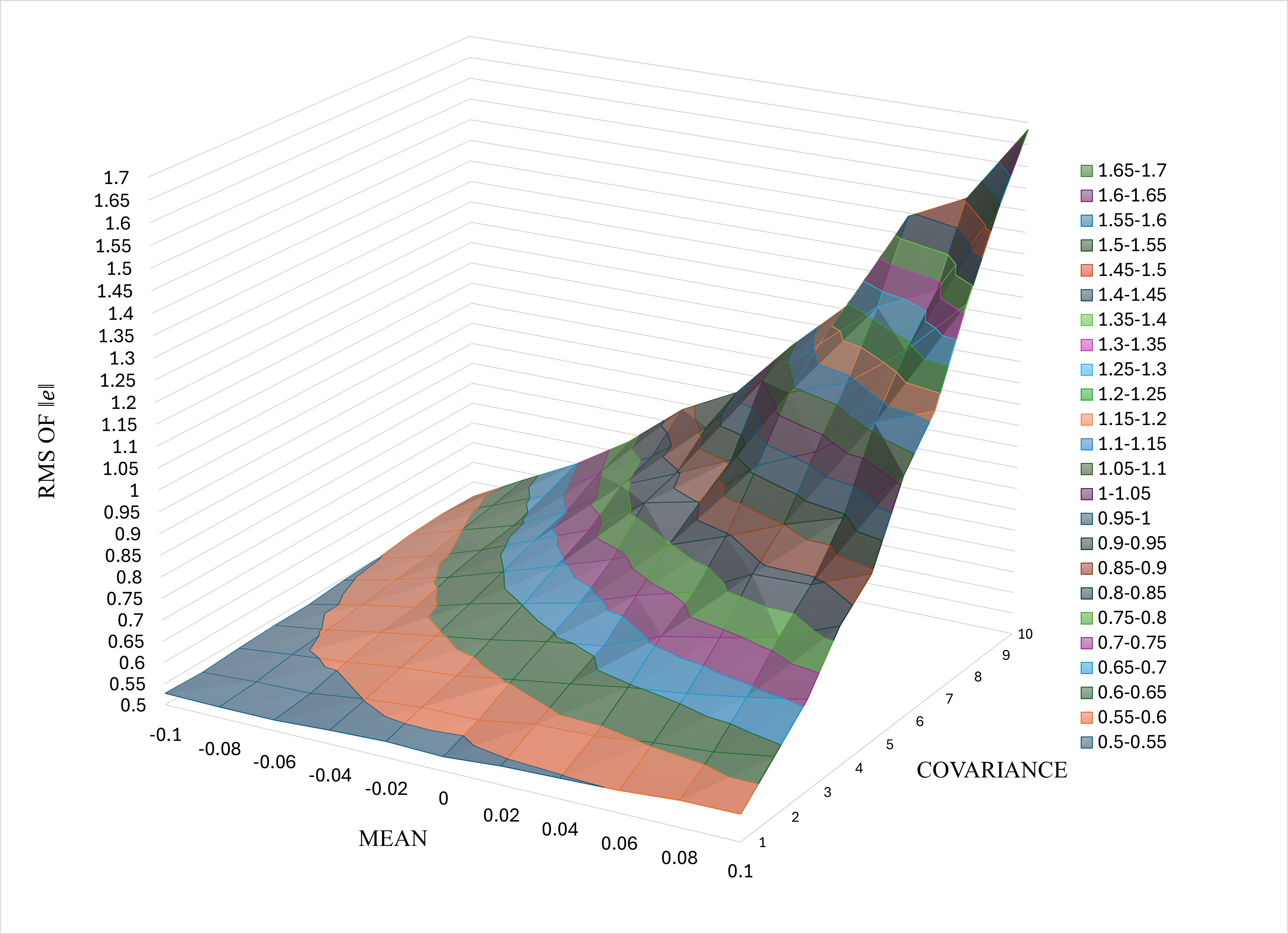}
\par\end{centering}
\centering{}\caption{\label{fig:mean-cov}Performance of the RMS of the tracking error
with respect to changes in mean and covariance of the stochastic noise
for the developed Lb-DNN controller.}
\end{figure}

Based on Figure \ref{fig:track. err.}, the tracking error demonstrates
an exponential convergence to its ultimate bound. The developed Lb-DNN
controller achieves a root-mean-square (RMS) tracking error norm of
$0.533$. Figure \ref{fig:mean-cov} illustrates the evolution of
the RMS of the norm of the tracking error in response to variations
in the mean and covariance of the stochastic noise. As shown in Figure
\ref{fig:mean-cov}, the RMS tracking error tends to increase as noise
mean and covariance grow. The smallest RMS tracking error of 0.524
occurs at a noise mean of -0.1 and a covariance of 2, whereas the
largest RMS tracking error of 1.684 is observed at a noise mean of
0.1 and a covariance of 10.

\section{Conclusion\label{sec: Conclusion}}

An Lb-DNN adaptive controller is developed for a stochastic process
modeled by a control-affine nonlinear stochastic differential equation
to achieve a trajectory tracking objective. Three Lb-DNNs are developed
to compensate for deterministic and non-deterministic uncertainties
within the closed-loop error system. The proposed Lb-DNN adaptive
controller and the stability-driven update laws ensure the tracking
error is uniformly ultimately bounded in probability, with a rigorous
probability analysis. The result is supported by a probability analysis
without the common assumption of vanishing noise and in the presence
of unstructured model uncertainty. Simulations are performed on a
nonlinear stochastic dynamical system to show the efficacy of the
proposed method.

\section*{Appendix}

\subsection*{Proof of Lemma \ref{thm:probability}}

Let the strong Markov process $\tilde{{\tt z}}\left(t\right)$ be
defined as
\begin{equation}
\text{\ensuremath{\tilde{{\tt z}}\left(t\right)\tq\begin{cases}
{\tt z}\left(t\right), & \text{for }t<\tau_{m},\\
0, & \text{for }t\geq\tau_{m}.
\end{cases}}}\label{eq: z tilde}
\end{equation}
Notice that if $t\geq\tau_{m}$, then $\left\Vert \tilde{{\tt z}}\left(t\right)\right\Vert =0$,
and if $t<\tau_{m}$, then $\left\Vert \tilde{{\tt z}}\left(t\right)\right\Vert =\left\Vert {\tt z}\left(t\right)\right\Vert $.
Therefore, since ${\tt V}$ is non-negative and ${\tt V}\left(0\right)=0$
(according to (A1)), ${\tt V\left(\tilde{{\tt z}}\left(t\right)\right)\leq{\tt V}}\left({\tt z}\left(t\right)\right)$.
Hence, the following relation holds for all $t\in\RR_{\geq0}$
\begin{equation}
{\rm E}\left[{\tt V}\left(\tilde{{\tt z}}\left(t\right)\right)\right]\leq{\rm E}\left[{\tt V}\left({\tt z}\left(t\right)\right)\right].\label{eq:Ebound1}
\end{equation}
The solution to stochastic differential inequality given by $\mathcal{L}{\tt V}\left({\tt z}\right)\leq-\kappa_{1}{\tt V}\left({\tt z}\right)+\kappa_{2}$
is (see \cite[Thm. 4.1]{Deng.Krstic.ea2001})
\begin{equation}
{\rm E}\left[{\tt V}\left({\tt z}\right)\right]\leq{\tt V}\left({\tt z}\left(0\right)\right)\exp\left(-\kappa_{1}t\right)+\frac{\kappa_{2}}{\kappa_{1}},\label{eq:solution to LV}
\end{equation}
for all $t\in\RR_{\geq0}$. Applying (\ref{eq:solution to LV}) on
(\ref{eq:Ebound1}) yields 
\begin{equation}
{\rm E}\left[{\tt V}\left(\tilde{{\tt z}}\right)\right]\leq{\tt V}\left({\tt z}\left(0\right)\right)\exp\left(-\kappa_{1}t\right)+\frac{\kappa_{2}}{\kappa_{1}},\label{eq: umm idk}
\end{equation}
for all $t\in\RR_{\geq0}$. 

The solution of $\mathcal{L}{\tt V}$ given in (\ref{eq:solution to LV})
shows that ${\rm E}\left[{\tt V}\left({\tt z}\right)\right]$ strictly
decreases until it reaches the bound $\frac{\kappa_{2}}{\kappa_{1}}$.
Once ${\rm E}\left[{\tt V}\left({\tt z}\right)\right]$ reaches this
bound, it stays bounded by $\frac{\kappa_{2}}{\kappa_{1}}$. However,
inside this bound, ${\rm E}\left[{\tt V}\left({\tt z}\right)\right]$
may increase, decrease, or stay constant. With this explanation provided,
different cases of behavior of ${\rm E}\left[{\tt V}\left({\tt z}\right)\right]$
are investigated here. To study the behavior of ${\rm E}\left[{\tt V}\left({\tt z}\right)\right]$
before reaching the ultimate bound, let $\tau_{B}\triangleq\inf\left\{ t\geq0:{\tt V}\left(\tilde{{\tt z}}(t)\right)\leq\frac{\kappa_{2}}{\kappa_{1}}\right\} $.
For $t\in\left[0,\tau_{B}\right)$, ${\tt V}$ is a supermartingale
(see \cite[Thm. C.4]{Oeksendal.Oeksendal2003}). Applying Doob's maximal
inequality (see \cite[Page 275]{LeGall2022}) and (\ref{eq: umm idk})
yields
\begin{gather}
{\rm P}\left(\underset{0\leq t\leq s<\tau_{B}}{\sup}{\tt V}\left(\tilde{{\tt z}}\left(s\right)\right)\geq\lambda\right)\leq\frac{1}{\lambda}{\rm E}\left[\underset{t\to\tau_{B}^{-}}{\lim}{\tt V}\left(\tilde{{\tt z}}\left(t\right)\right)\right]\nonumber \\
\leq\frac{1}{\lambda}{\rm E}\left[{\tt V}\left(\tilde{{\tt z}}\left(t\right)\right)\right]\leq\frac{1}{\lambda}\left({\tt V}\left({\tt z}\left(0\right)\right)\exp\left(-\kappa_{1}t\right)+\frac{\kappa_{2}}{\kappa_{1}}\right).\label{eq: supermartingale}
\end{gather}
To study the behavior of ${\rm E}\left[{\tt V}\left({\tt z}\right)\right]$
after reaching the ultimate bound, consider the interval $\mathcal{I}=\left[\tau_{B},\infty\right)$.
Without loss of generality, partition this interval into subintervals,
defining $\mathcal{I}=\left[\tau_{B},\tau_{1}\right)\cup\left[\tau_{1},\tau_{2}\right)\cup\left[\tau_{2},\tau_{3}\right)\cup\left[\tau_{3},\infty\right)$,
where each subinterval corresponds to a specific behavior of the process
${\tt V}$: on $\left[\tau_{B},\tau_{1}\right)$, ${\tt V}$ is a
supermartingale; on $\text{\ensuremath{\left[\tau_{1},\tau_{2}\right)}}$,
a submartingale; and on $\left[\tau_{2},\tau_{3}\right)$, a martingale.

For $t\in\left[\tau_{B},\tau_{1}\right)$, since ${\tt V}$ is a supermartingale,
using Doob's maximal inequality and (\ref{eq: umm idk}) yields
\begin{gather}
{\rm P}\left(\underset{\tau_{B}\leq t\leq s<\tau_{1}}{\sup}{\tt V}\left(\tilde{{\tt z}}\left(s\right)\right)\geq\lambda\right)\leq\frac{1}{\lambda}{\rm E}\left[\underset{t\to\tau_{1}^{-}}{\lim}{\tt V}\left(\tilde{{\tt z}}\left(t\right)\right)\right]\nonumber \\
\leq\frac{1}{\lambda}{\rm E}\left[{\tt V}\left(\tilde{{\tt z}}\left(t\right)\right)\right]\leq\frac{1}{\lambda}\left({\tt V}\left({\tt z}\left(0\right)\right)\exp\left(-\kappa_{1}t\right)+\frac{\kappa_{2}}{\kappa_{1}}\right).\label{eq: supermartingale-1}
\end{gather}
For $t\in\left[\tau_{1},\tau_{2}\right)$, since ${\tt V}$ is a submartingale
(see \cite[Thm. C.4]{Oeksendal.Oeksendal2003}), using Doob's maximal
inequality yields
\begin{gather}
{\rm P}\left(\underset{\tau_{1}\leq t\leq s<\tau_{2}}{\sup}{\tt V}\left(\tilde{{\tt z}}\left(s\right)\right)\geq\lambda\right)\leq\frac{1}{\lambda}{\rm E}\left[\lim_{t\to\tau_{2}^{-}}{\tt V}\left(\tilde{{\tt z}}\left(t\right)\right)\right]\nonumber \\
\leq\frac{1}{\lambda}\frac{\kappa_{2}}{\kappa_{1}}\leq\frac{1}{\lambda}\left({\tt V}\left({\tt z}\left(0\right)\right)\exp\left(-\kappa_{1}t\right)+\frac{\kappa_{2}}{\kappa_{1}}\right).\label{eq:submartingale}
\end{gather}
For $t\in\left[\tau_{2},\tau_{3}\right)$, since ${\tt V}$ is a martingale,
using Doob's maximal inequality and (\ref{eq: umm idk}) yields
\begin{gather}
{\rm P}\left(\underset{\tau_{2}\leq t\leq s<\tau_{3}}{\sup}{\tt V}\left(\tilde{{\tt z}}\left(s\right)\right)\geq\lambda\right)\leq\frac{1}{\lambda}{\rm E}\left[\underset{t\to\tau_{3}^{-}}{\lim}{\tt V}\left(\tilde{{\tt z}}\left(t\right)\right)\right]\nonumber \\
\leq\frac{1}{\lambda}{\rm E}\left[{\tt V}\left(\tilde{{\tt z}}\left(t\right)\right)\right]\leq\frac{1}{\lambda}\left({\tt V}\left({\tt z}\left(0\right)\right)\exp\left(-\kappa_{1}t\right)+\frac{\kappa_{2}}{\kappa_{1}}\right).\label{eq: supermartingale-1-1}
\end{gather}
Since $\left[0,\infty\right)=\left[0,\tau_{B}\right)\cup\mathcal{I}$,
and the bounds over each time interval are identical, no generality
is lost. Therefore, using (\ref{eq: supermartingale})-(\ref{eq: supermartingale-1-1})
yields{\small{}
\begin{equation}
{\rm P}\left(\underset{t\leq s<\infty}{\sup}{\tt V}\left(\tilde{{\tt z}}\left(s\right)\right)\geq\lambda\right)\leq\frac{1}{\lambda}\left({\tt V}\left({\tt z}\left(0\right)\right)\exp\left(-\kappa_{1}t\right)+\frac{\kappa_{2}}{\kappa_{1}}\right).\label{eq: OKKK}
\end{equation}
}Note that for $t\geq\tau_{m}$, ${\tt V}\left(\tilde{{\tt z}}\left(t\right)\right)\neq{\tt V}\left({\tt z}\left(t\right)\right)$.
Then, it holds that{\footnotesize{}
\begin{gather}
{\rm P}\left(\underset{t\leq s<\infty}{\sup}\left|{\tt V}\left(\tilde{{\tt z}}\left(s\right)\right)-{\tt V}\left({\tt z}\left(s\right)\right)\right|\geq\lambda\right)\nonumber \\
\leq{\rm P}\left(\underset{0\leq s<\infty}{\sup}\left|{\tt V}\left(\tilde{{\tt z}}\left(s\right)\right)-{\tt V}\left({\tt z}\left(s\right)\right)\right|\geq\lambda\right).\label{eq: abs of probability -1}
\end{gather}
}For $t\leq\tau_{m}$, where ${\tt V}\left(\tilde{{\tt z}}\left(t\right)\right)={\tt V}\left({\tt z}\left(t\right)\right)$,
it follows that ${\rm P}\left(\underset{t\leq s<\infty}{\sup}{\tt V}\left(\tilde{{\tt z}}\left(s\right)\right)\geq\lambda\right)={\rm P}\left(\underset{t\leq s<\infty}{\sup}{\tt V}\left({\tt z}\left(s\right)\right)\geq\lambda\right)$.
Therefore,{\scriptsize{}
\begin{gather}
\left|{\rm P}\left(\underset{t\leq s<\infty}{\sup}{\tt V}\left(\tilde{{\tt z}}\left(s\right)\right)\geq\lambda\right)-{\rm P}\left(\underset{t\leq s<\infty}{\sup}{\tt V}\left({\tt z}\left(s\right)\right)\geq\lambda\right)\right|\nonumber \\
={\rm P}\left(\underset{t\leq s<\infty}{\sup}\left|{\tt V}\left(\tilde{{\tt z}}\left(s\right)\right)-{\tt V}\left({\tt z}\left(s\right)\right)\right|\geq\lambda\right)=0.\label{eq: abs of probability}
\end{gather}
}Similarly, when $t>\tau_{m}$, ${\tt V}\left(\tilde{{\tt z}}\left(t\right)\right)=0$,
which implies ${\rm P}\left(\underset{t\leq s<\infty}{\sup}{\tt V}\left(\tilde{{\tt z}}\left(s\right)\right)\geq\lambda\right)=0$,
ensuring that (\ref{eq: abs of probability}) holds for all $t\in\RR_{\geq0}$.
Applying the triangle inequality on the left side of (\ref{eq: abs of probability})
yields{\scriptsize{}
\begin{gather}
{\rm P}\left(\underset{t\leq s<\infty}{\sup}{\tt V}\left({\tt z}\left(s\right)\right)\geq\lambda\right)-{\rm P}\left(\underset{t\leq s<\infty}{\sup}{\tt V}\left(\tilde{{\tt z}}\left(s\right)\right)\geq\lambda\right)\nonumber \\
\leq\left|{\rm P}\left(\underset{t\leq s<\infty}{\sup}{\tt V}\left(\tilde{{\tt z}}\left(s\right)\right)\geq\lambda\right)-{\rm P}\left(\underset{t\leq s<\infty}{\sup}{\tt V}\left({\tt z}\left(s\right)\right)\geq\lambda\right)\right|.\label{eq: abs of probability 1}
\end{gather}
}Solving for ${\rm P}\left(\underset{t\leq s<\infty}{\sup}{\tt V}\left({\tt z}\left(s\right)\right)\geq\lambda\right)$
in (\ref{eq: abs of probability 1}), and using (\ref{eq: abs of probability -1})
and (\ref{eq: abs of probability}) yields{\scriptsize{}
\begin{gather}
{\rm P}\left(\underset{t\leq s<\infty}{\sup}{\tt V}\left({\tt z}\left(s\right)\right)\geq\lambda\right)\leq{\rm P}\left(\underset{0\leq s<\infty}{\sup}\left|{\tt V}\left(\tilde{{\tt z}}\left(s\right)\right)-{\tt V}\left({\tt z}\left(s\right)\right)\right|\geq\lambda\right)\nonumber \\
+{\rm P}\left(\underset{t\leq s<\infty}{\sup}{\tt V}\left(\tilde{{\tt z}}\left(s\right)\right)\geq\lambda\right).\label{eq: close to the final step}
\end{gather}
}Using Markov's inequality (see \cite[Page 153]{LeGall2022}), it
can be stated that{\footnotesize{}
\begin{gather}
{\rm P}\left(\underset{0\leq s<\infty}{\sup}\left|{\tt V}\left(\tilde{{\tt z}}\left(s\right)\right)-{\tt V}\left({\tt z}\left(s\right)\right)\right|\geq\lambda\right)\nonumber \\
={\rm P}\left(\underset{0\leq s<\infty}{\sup}{\tt V}\left({\tt z}\left(s\right)\right)\geq m\right)\leq\frac{1}{m}{\tt V}\left({\tt z}\left(0\right)\right).\label{eq: basically final}
\end{gather}
}Therefore, using (\ref{eq: OKKK}), (\ref{eq: close to the final step}),
and (\ref{eq: basically final}) yields{\small{}
\begin{gather}
{\rm P}\left(\underset{t\leq s<\infty}{\sup}{\tt V}\left({\tt z}\left(s\right)\right)\geq\lambda\right)\leq\frac{1}{m}{\tt V}\left({\tt z}\left(0\right)\right)\nonumber \\
+\frac{1}{\lambda}{\tt V}\left({\tt z}\left(0\right)\right)\exp\left(-\kappa_{1}t\right)+\frac{\kappa_{2}}{\kappa_{1}\lambda}.\label{eq:z}
\end{gather}
}{\small\par}

{\small{}}{\small\par}

\bibliographystyle{ieeetr}
\bibliography{2C__Users_saied_AKBARIS_bibtex_bib_ncrbibs_encr,3C__Users_saied_AKBARIS_bibtex_bib_ncrbibs_master,4C__Users_saied_AKBARIS_bibtex_bib_ncrbibs_ncr}

\begin{IEEEbiography}[{\includegraphics[scale=0.09]{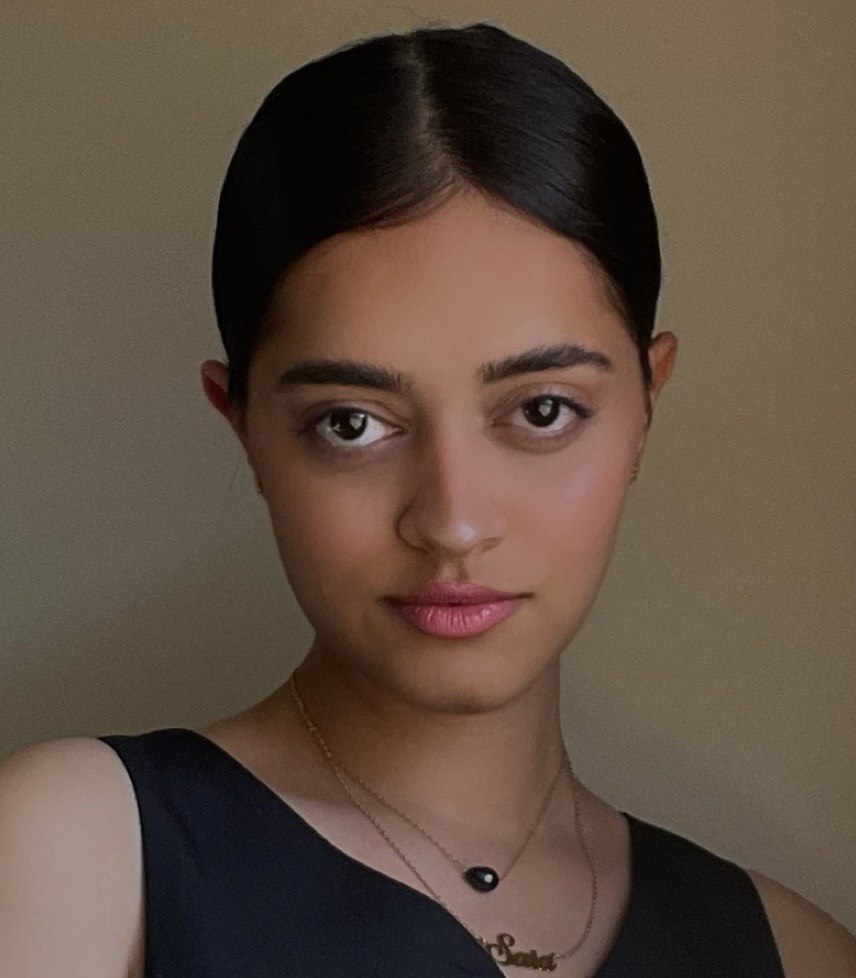}}]{Saiedeh Akbari}
 is a Ph.D. candidate in the department of Mechanical and Aerospace
Engineering at University of Florida. Her research focuses on developing
adaptive learning-based control strategies for stochastic nonlinear
systems. Saiedeh completed her Bachelor of Science in Mechanical Engineering
at KN Toosi University of Technology in 2020. During her undergraduate
studies, she conducted research on discrete-time sliding mode control
for permanent magnet DC motors. She then received her Master of Science
in Mechanical Engineering at The University of Alabama in 2023, where
she worked on nonlinear adaptive control techniques for hybrid systems,
with applications in rehabilitation robotics.
\end{IEEEbiography}

\begin{IEEEbiography}[{\includegraphics[scale=0.35]{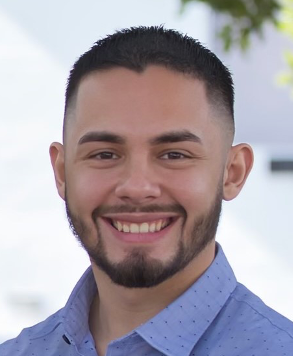}}]{Cristian F. Nino}
 received the B.S. degrees in Mathematics and Mechanical Engineering
from the University of Florida, Gainesville, FL, USA. He subsequently
earned the M.S. degree in Mechanical Engineering, also from the University
of Florida, with a focus on control systems. He is currently pursuing
the Ph.D. degree in Mechanical Engineering at the University of Florida
under the guidance of Dr. Warren Dixon. Mr. Nino is a recipient of
the SMART Scholarship, the NSF Fellowship, and the Machen Florida
Opportunity Scholarship. His research interests include robust adaptive
nonlinear control, multi-agent target tracking, distributed state
estimation, reinforcement learning, Lyapunov-based deep learning,
and geometric mechanics and control.
\end{IEEEbiography}

\begin{IEEEbiography}[{\includegraphics[scale=0.105]{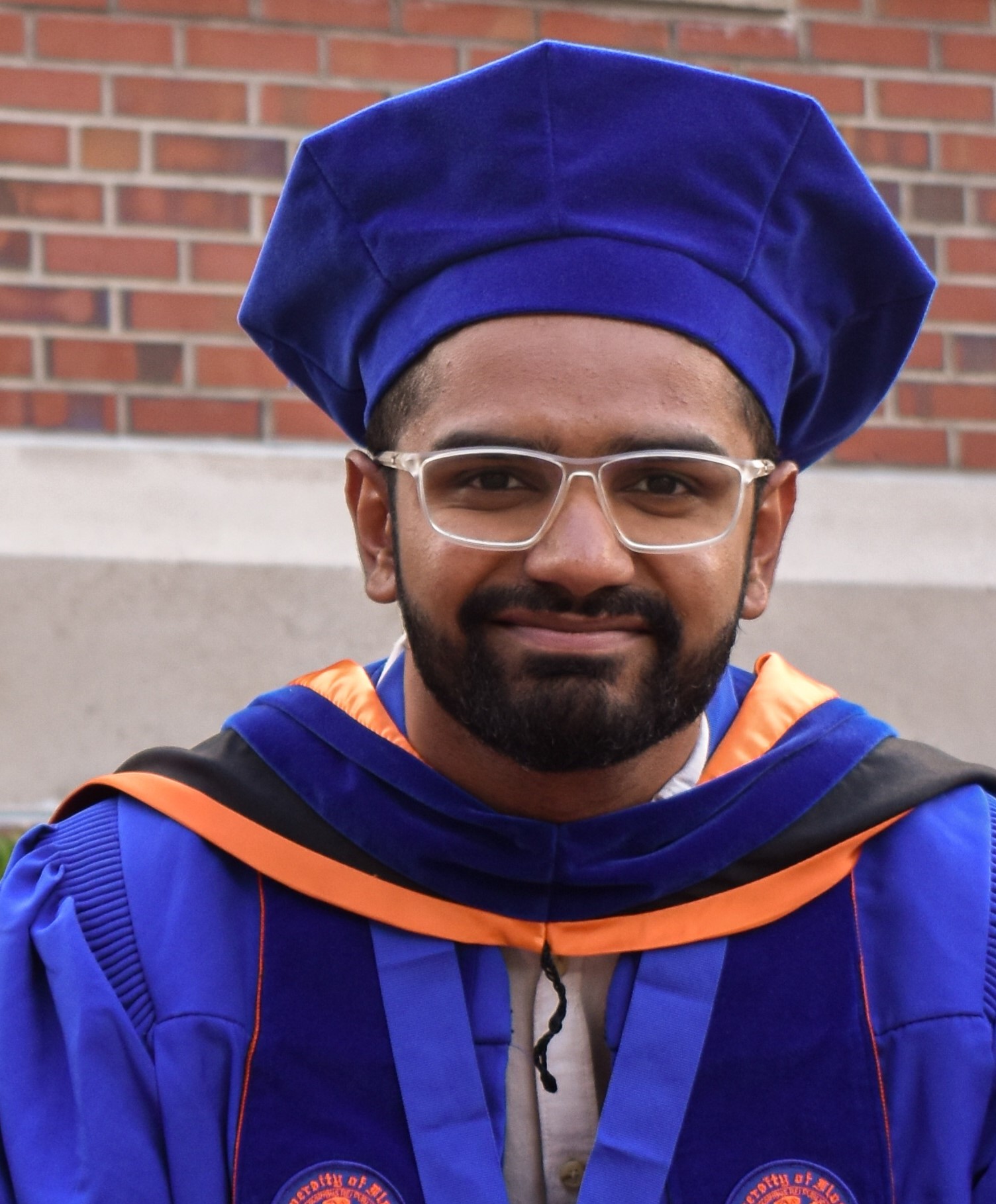}}]{Omkar Sudhir Patil}
 received his Bachelor of Technology (B.Tech.) degree in production
and industrial engineering from Indian Institute of Technology (IIT)
Delhi in 2018, where he was honored with the BOSS award for his outstanding
bachelor's thesis project. In 2019, he joined the Nonlinear Controls
and Robotics (NCR) Laboratory at the University of Florida under the
guidance of Dr. Warren Dixon to pursue his doctoral studies. Omkar
received his Master of Science (M.S.) degree in mechanical engineering
in 2022 and Ph.D. in mechanical engineering in 2023 from the University
of Florida. During his PhD studies, he was awarded the Graduate Student
Research Award for outstanding research. In 2023, he started working
as a postdoctoral research associate at NCR Laboratory, University
of Florida. His research focuses on the development and application
of innovative Lyapunov-based nonlinear, robust, and adaptive control
techniques. 
\end{IEEEbiography}

\begin{IEEEbiography}[{\includegraphics[scale=0.0625]{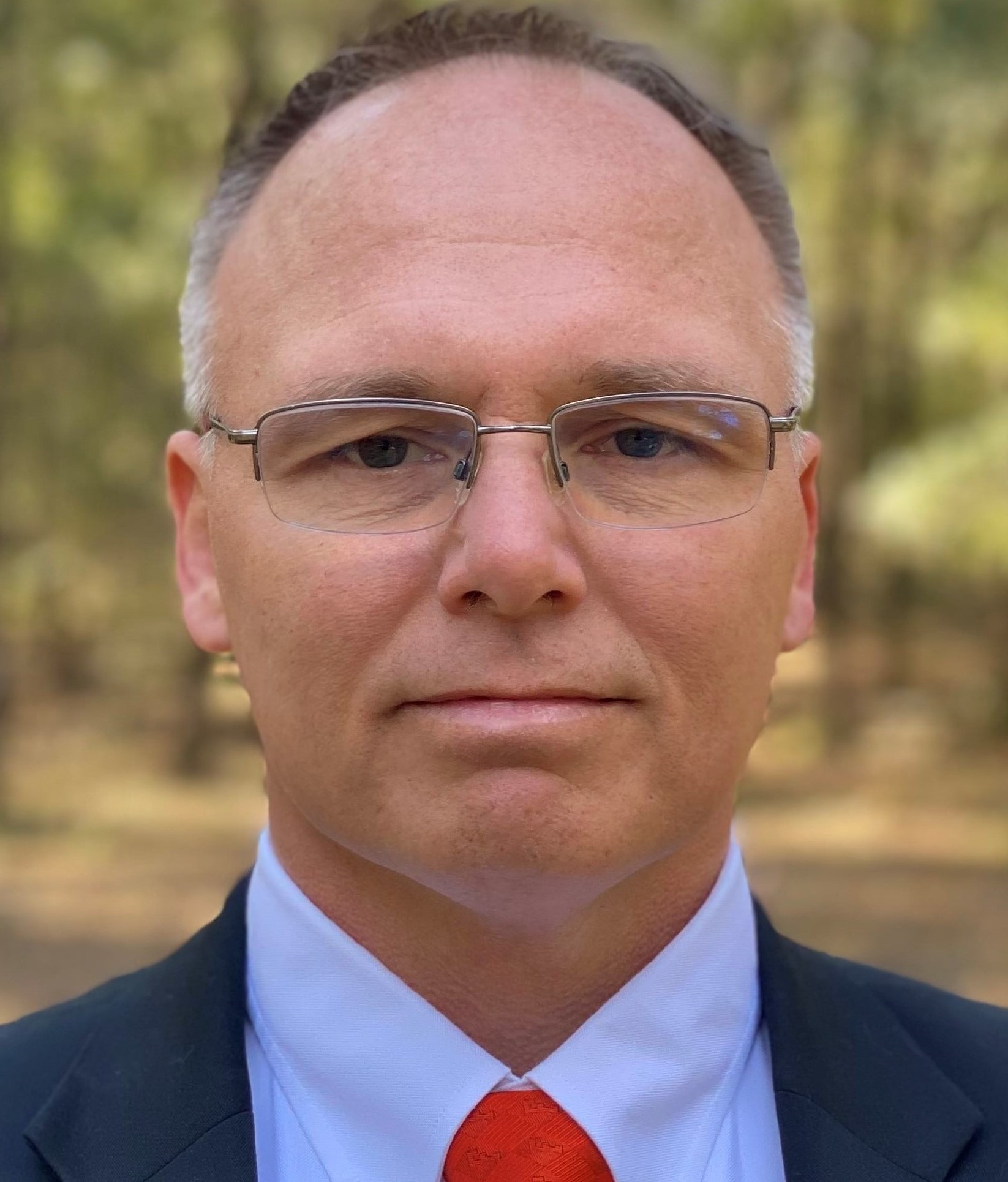}}]{Prof. Warren Dixon}
 received his Ph.D. in 2000 from the Department of Electrical and
Computer Engineering from Clemson University. He worked as a research
staff member and Eugene P. Wigner Fellow at Oak Ridge National Laboratory
(ORNL) until 2004, when he joined the University of Florida in the
Mechanical and Aerospace Engineering Department. His main research
interest has been the development and application of Lyapunov-based
control techniques for uncertain nonlinear systems. His work has been
recognized by the 2019 IEEE Control Systems Technology Award, (2017-2018
\& 2012-2013) University of Florida College of Engineering Doctoral
Dissertation Mentoring Award, 2015 \& 2009 American Automatic Control
Council (AACC) O. Hugo Schuck (Best Paper) Award, the 2013 Fred Ellersick
Award for Best Overall MILCOM Paper, the 2011 American Society of
Mechanical Engineers (ASME) Dynamics Systems and Control Division
Outstanding Young Investigator Award, the 2006 IEEE Robotics and Automation
Society (RAS) Early Academic Career Award, an NSF CAREER Award (2006-2011),
the 2004 Department of Energy Outstanding Mentor Award, and the 2001
ORNL Early Career Award for Engineering Achievement. He is an ASME
Fellow (2016) and IEEE Fellow (2016), was an IEEE Control Systems
Society (CSS) Distinguished Lecturer (2013-2018), served as the Director
of Operations for the Executive Committee of the IEEE CSS Board of
Governors (BOG) (2012-2015), and served as an elected member of the
IEEE CSS BOG (2019-2020). His technical contributions and service
to the IEEE CSS were recognizd by the IEEE CSS Distinguished Member
Award (2020). He was awarded the Air Force Commander's Public Service
Award (2016) for his contributions to the U.S. Air Force Science Advisory
Board. He is currently or formerly an associate editor for ASME Journal
of Journal of Dynamic Systems, Measurement and Control, Automatica,
IEEE Control Systems, IEEE Transactions on Systems Man and Cybernetics:
Part B Cybernetics, and the International Journal of Robust and Nonlinear
Control.
\end{IEEEbiography}

\end{document}